# Toe-Heal-Air-Injection Thermal Recovery Production Prediction and Modelling Using Quadratic Poisson Polynomial Regression

Alan Rezazadeh, Applied Research and Innovation Services
Southern Alberta Institute of Technology, Calgary, Alberta, Canada

## 1  Abstract

This research paper explores application of multivariable regression models using only reservoir temperatures for predicting oil and gas production in a Toe-Heal-Air-Injection (THAI) enhanced oil recovery process. This paper discusses effects of statistical interaction between thermocouples by using second degree quadratic polynomials, which showed significant production forecast accuracy. Interactions among thermocouples statistically include temperature of larger reservoir areas, hence improving the predictive models. The interaction of two thermocouples can be interpreted as temperature gradient of combustion zone as moving forward during THAI operations life cycle.

Second degree polynomial regression including interactions showed major prediction improvement for both oil and natural gas productions compare to simple regression models. Application of Poisson regression slightly improved prediction accuracy for oil production and was less effective on improving natural gas production predictions. Quadratic Poisson regression models showed realistic production prediction method for both oil and gas production values, due to the nature of Poisson probability distribution which is non-negative for rates and count values.

## 2  Introduction

The main objective of this research paper is to present a statistical parametric approach building regression models for predicting oil and natural gas production in THAI (Toe Heel Air Injection) methods. This paper discusses two different regression models (1$^{st}$ and 2$^{nd}$) degree polynomials among with two different statistical distribution (Poisson and Normal) to build the predictive models.

The parametric regression methods explored in this paper, simulate the reservoir behaviour with mathematical models such as Normal or Poisson Regressions, distinguished from non-parametric methods (i.e. Neural Network and Deep Learning). The parametric methods, simulate the observed data points using predefined mathematical formulas such as polynomials and thermocouple temperature readings as predictors. This approach is successful when the modelled process can be mathematically described by the regression algorithms. The advantage of parametric models are usually faster estimations, less amount of data required to train the models, higher prediction rates specifically for process conditions without captured data to train the models.

The results presented in this paper show conclusively using second degree quadratic polynomials offering the best predictive models, due to inclusion of interaction between thermocouples (i.e. temperature) within different sections of reservoir. Adding Poisson distribution increased the predictability of second order polynomials while making the models more realistic which is due to nature of Poisson distribution that is mostly used for count, or rate values.

The focus of this paper is on building prediction models based on generalized regression, using second degree polynomials, incorporating interactions between predictors as well as comparisons of Normal and Poisson distributions, which improved the predictability of the models. This paper explores the relationships between temperature observations and production values, independent from other factors such as air injection volumes, oil viscosity, and production of Hydrogen, Oxygen, Nitrogen, Carbon Monoxide, Carbon Dioxide, and Methane. A more complete modelling of production based on temperature readings among with other characteristics listed above offer more in-depth modelling and statistical analysis opportunity.

## 3  Kerrobert Toe Heel Air Injection Study Area

The study area is located in Saskatchewan, Canada, and targeted Waseca formation within Mannvile sandstone pool. Contrary to conventional fireflood method of utilizing two vertical wells for air injection and production, THAI process is based on a vertical well for air injection and a horizontal production well. The project comprised of 12 well pairs, each consist of a production and injection well as illustrated in Figure 1. This experiment of thermal recovery method is based on injecting air into the reservoir by a vertical well, igniting and consequently combusting the immobile heavy oil, increasing the reservoir temperature, hence thermally cracking and upgrading the mobilized oil to be drained by the horizontal production well. During the THAI process, approximately about 5% - 10% of the Original Oil In Place (OOIP) is combusted, and the process is managed by controlling the amount of air injected [3]. The reservoir temperature increase also positively affects the production by decreasing viscosity of the mobilized oil and leaving a drained burned reservoir behind (Figure 1).

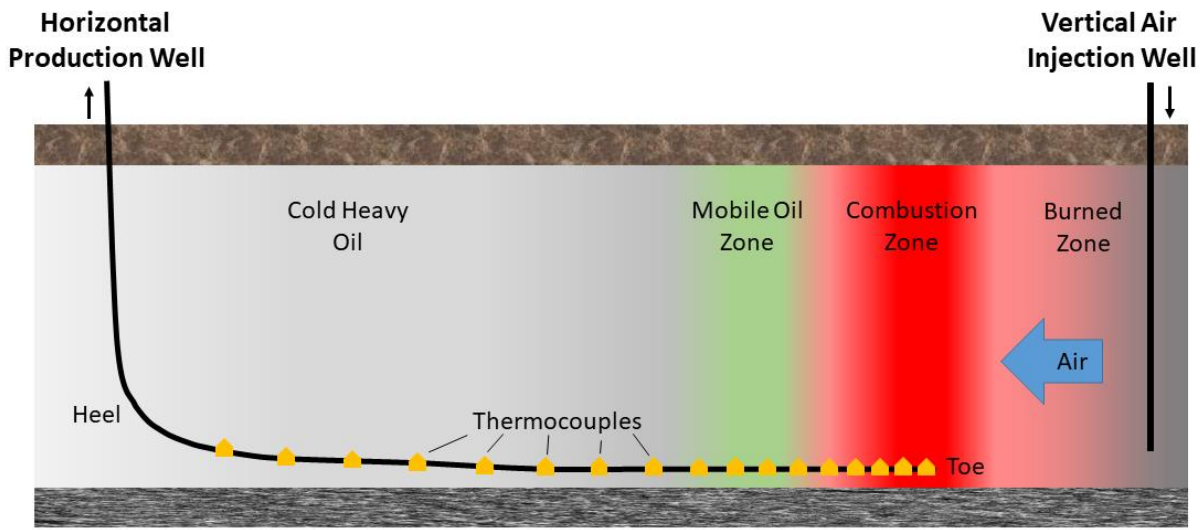

Figure 1. Configuration for each well pair, comprised of vertical air injection and horizontal production wells.

For this project, well pairs 1 to 7 were drilled within the east pad and five wells 8 to 12 within the west pad. Well pairs 1 and 2 started operations as pilot testing in 2009 and other well pairs commenced production in 2011 [1]. Well pairs 4 and 12 only produced for a short period of time and due to sand production were abandoned, therefore, the data for these two wells were excluded from this analysis. Wells 10 and 12 air injections were commingled throughout the operations. The data capture for this operations started from November 2009, to May 2018, excluding the two year period starting at 2015, which the operations stopped and wells were shut-in due to managerial changes until late 2017.

## 4  Machine Learning Based on Regression Techniques

There are different methods for machine learning, depending to the objectives, data availability and prior knowledge of the process to be predicted (Figure 2). The objective of this paper is to use parametric methods for modelling the production values, based on temperature readings of the thermocouples. Parametric models use a predefined mathematical model to simulate the process behaviour. For example in this paper two estimation methods, first and second degree (quadratic) polynomials have been used for modelling the production of oil and gas. Meanwhile for each estimation method, two response distributions (i.e. Normal and Poisson) have been tested against the actual production data, errors of estimation are measured and analysed to select the most effective method.

Non-parametric models (i.e. Neural Networks) present a family of process modelling methods mostly used where the process behaviour is not conforming to any of the known predefined statistical models or there are no prior knowledge of the process behaviour available, which also require larger amount of training data and typically are more time consuming for building the models.

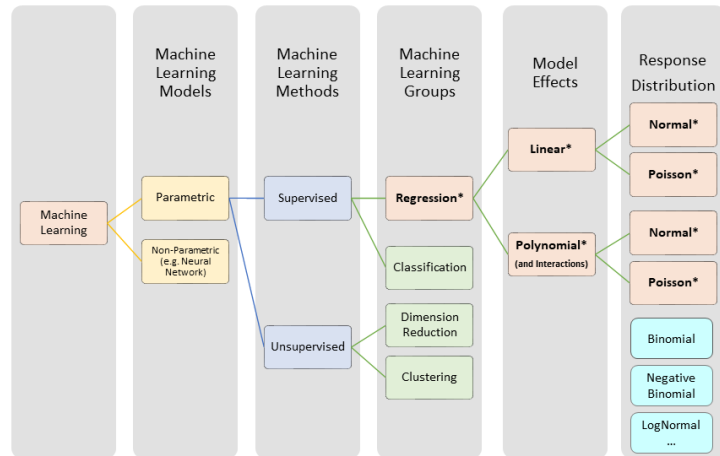

Figure 2. An oversimplified hierarchy of the machine learning algorithms.
(* algorithms used within this paper)

Regression analysis is a parametric learning framework to build models describing relationships between response (e.g. dependent variables) and predictors (e.g. independent variables). Regression analysis is based on parametric statistics that assumes the models can be built on one or more of predefined probability distributions such as Normal, Poisson or Gamma distributions, while the relationship between predictors and response follow a specific mathematical formulae defined by the regression model.

In this paper two methods of linear and second degree polynomial regressions among with Normal and Poisson distributions have been used to model the THAI process based on thermocouple temperature readings. In addition of the two probability distributions, the models include second degree quadratic polynomials and interactions between thermocouples to improve overall prediction rates as well.

The Regression analysis framework is based on dependent variable Y|X, where $Y = f(X) + \varepsilon$, and ε is the residual error, the difference between actual values of Y and calculated value of f(x). The objective is to have mean value of zero and Normal distribution for ε values, while minimizing the total value of residual error.

## 4.1    Normal Distribution

Normal Distribution, also known as Gaussian is a probability density function, used in wide range of applications, such as natural and social sciences. Normal distribution is a symmetric distribution function where most of observations are concentrated around the mean value and the probabilities for the values further away from the mean are reduced in both directions. The following formula describes the probability density function $\phi(x)$ of Normal distribution mathematically:

$$\phi(x) = \frac{1}{\delta\sqrt{2\pi}} e^{-\frac{1}{2}(\frac{x-\mu}{\delta})^2}$$

where, $\delta$ is the standard deviation, $\delta^2$ is variance and mean is $\mu$.

### 4.2 Poisson Distribution

Poisson distribution, named after Simeon Denis Poisson, French mathematician, presents the probability of occurring an event in a specified time interval or allocated space, given mean is known and time is an independent factor. Given mean shown as $\lambda$ for events or rates of a specified event over a constant time or space is known. In this study $\lambda$ is the mean production fluids or natural gas in square meter ($m^3$) per day, during the wells' life cycle. The following formula describes the probability density of Poisson function, for a specified mean $\lambda$, equals to variance

$$P(Y = y|\lambda) = \frac{\lambda^y e^{-\lambda}}{y!}$$

where $P(Y = y|\lambda)$ is the probability mass function for a specific count or rate $y$ given the mean $\lambda$.

For this analysis the objectives are using specific temperature readings vector $X = (x_1, \ldots, x_p)$ and Poisson distribution, predicting the most probable production, which is $\lambda$ the mean using and p number of regression parameters. Using Poisson regression for predicting mean daily production $\lambda$ using the following formulae:

$$\lambda = e^{X\beta}$$

where X is the predictor's vector and $\beta$ is the regression coefficients vector. One of the major advantages of using Poisson regression is calculated $\lambda$ value is non-negative and can be considered as estimated production rate [8].

The main constraint with Poisson distribution is definition of mean ($\lambda$) to be equal to variance. In most engineering applications the variance is larger than mean which is called over-dispersion. Similarly, if variance is less than mean the situation will be called under-dispersion. Poisson distribution assumes $\lambda$, the mean is small and large values for $\lambda$ are uncommon [7].

## 5 Linear Regression with Normal Distribution

The regression models discussed in this paper are members of Generalized Linear Models (GLM) family, allowing combination of different model effects (polynomial degree and interactions) and distributions (Normal vs. Poisson). Although GLM supports wide range of combinations for effects and distributions, only a subset of selections were used in this paper that provided higher success prediction rates.

### 5.1 Univariable (Simple) Linear Regression

Simple linear regression is a statistical method of finding relationship between two continuous parameters, where one parameter denoted x called predictor (explanatory, or independent) and the other parameter denoted y is called response (outcome or dependent) variable. The following is a simple linear regression, with only one predictor x, and $i$ is the number of the data observations:

$$\hat{y}_i = \beta_0 + \beta_1 x_i$$
$$y_i = \beta_0 + \beta_1 x_i + \varepsilon_i$$
$$\varepsilon_i = y_i - \hat{y}_i$$

where $\varepsilon_i$ residual error, is the difference between actual observation ($y_i$) and predicted value ($\hat{y}_i$), which is desired to be minimized.

## 5.2 *Multivariable Linear Regression*

Multivariable linear regression also known as multiple linear regression consists of two or more predictors, building an approximation formulae for predicting the response variable *y*. For instance the following is a multivariable linear regression using two predictors:

$$y_i = \beta_0 + \beta_1 x_{i1} + \beta_2 x_{i2} + \varepsilon_i$$

where $\beta_j$ are the coeffiecients and $x_{ij}$ (e.g. $x_{i1}$ and $x_{i2}$) are the predictors, and *i* specifies the data point.

Expanding the relationship from two to *k* predictors the multivariable regression formulae can be written as:

$$y_i = \beta_0 + \beta_1 x_{i1} + \cdots + \beta_k x_{ik} + \varepsilon_i = \beta_0 + \sum_{j=1}^{k} \beta_j x_{ij} + \varepsilon_i \quad (1)$$

where, $y_i$ is the response variable, $\beta_0$ is intercept, and $\varepsilon_i$ is residual error for data point *i*. Formula (1) has been used in this paper for multivariable regression model.

## 5.3 *Univariable Quadratic Polynomial Regression*

Using polynomial regression, the value of response variable *y* is estimated by a quadratic formula, using a predictor variable *x*:

$$y = \beta_0 + \beta_1 x + \beta_2 x^2 + \varepsilon.$$

Expanding the formula to multiple observations by the following relationship, where *i* indicates the data point observation:

$$y_i = \beta_0 + \beta_1 x_i + \beta_2 x_i^2 + \varepsilon_i.$$

Above formula describes response variable $y_i$ based on readings of one predictor variable $x_i$.

## 5.4 *Multivariable Quadratic Polynomial Regression*

Multivariable Quadratic Regression is a powerful data mining method of establishing relationships based on second degree quadratic polynomials. Application of this model implies the response variable *y* has linear relationship with each predictor, their interactions with each other, as well as the predictor's squared value.

The formulas for a two predictor quadratic polynomial can be written as the following, where $x_1$ and $x_2$ are the variables [1]:

$$y = \beta_0 + \beta_1 x_1 + \beta_2 x_2 + \beta_{11} x_1^2 + \beta_{12} x_1 x_2 + \beta_{22} x_2^2 + \varepsilon.$$

Expanding the formula to include the data readings results the following relationship:

$$y_i = \beta_0 + \beta_1 x_{i1} + \beta_2 x_{i2} + \beta_{11} x_{i1}^2 + \beta_{12} x_{i1} x_{i2} + \beta_{22} x_{i2}^2 + \varepsilon_i.$$

Generalizing the relationship above to include *p* predictors, results the following formula:

$$y_i = \beta_0 + \sum_{j=1}^{p} \beta_j x_{ij} + \sum_{j=1}^{p}\sum_{k<j}^{p} \beta_{jk} x_{ij} x_{ik} + \sum_{j=1}^{p} \beta_{jj} x_{ij}^2 + \varepsilon_i.$$

Formula above contains multiplication of the two predictors $(x_j x_k)$, which are the interactions in the regression, which are also known as second degree partial factorial regression (two-way interactions). Expanding formula above to include centralization by subtracting mean from predictors in interactions and squared values result formula (2):

$$y_i = \beta_0 + \sum_{j=1}^{p} \beta_j x_{ij} + \sum_{j=1}^{p}\sum_{k<j}^{p} \beta_{jk} (x_{ij} - \mu_j)(x_{ik} - \mu_k) + \sum_{j=1}^{p} \beta_j (x_{ij} - \mu_j)^2 + \varepsilon_i \quad (2)$$

where $\mu_j$ is the mean for predictor $x_j$.

Centering of predictors are implemented at polynomial and interaction predictors to reduce collinearity between model effects [10]. Formulae (2) is implemented by the software (JMP® Pro 15 from SAS Institute Inc.) used for this analysis and the results displayed throughout this paper are based on centering predictors for interactions and second degree polynomials [5].

## 6 Linear Regression with Poisson Distribution

The linear regressions described earlier are based on Normal distribution which is used for continuous predictors and response variables. The formulas (1) and (2) imply the assumption that distribution of response variable are Normal. In this section the assumption for analysis is the response variable follows Poisson distribution, which is used for discrete probability functions. Although Poisson distribution is defined as discrete probability function, Generalized Linear Models does not require discrete response for application of Poisson distribution [11].

### 6.1 Multivariable Poisson Regression

The regression formulas based on the Poisson distribution for a multivariable (first degree) regression will be the following:

$$y_i = e^{\beta_0 + \sum_{j=1}^{k} \beta_j x_{ij}} + \varepsilon_i \quad (3)$$

where the exponent component is same as a simple linear regression with Normal distribution and $\varepsilon_i$ is the residual error. Formulae (3) has been used in the analysis for first degree multivariable Poisson regressions.

### 6.2 Multivariable Quadratic Polynomial Poisson Regression

Similarly the second degree polynomial regression including the interactions (two-way) using Poisson distribution can be written as the following:

$$y_i = e^{\beta_0 + \sum_{j=1}^{p} \beta_j x_{ij} + \sum_{j=1}^{p}\sum_{k<j}^{p} \beta_{jk}(x_{ij}-\mu_j)(x_{ik}-\mu_k) + \sum_{j=1}^{p} \beta_j(x_{ij}-\mu_j)^2} + \varepsilon_i. \quad (4)$$

Formulae (4) has been used in the analysis which also includes centering for interactions and squared predictors values. Centering is used for reducing multi-collinearity effects in the interactions and squared predictors [5].

### 6.3 Model Comparisons

In this report multiple regression model using different two different regression effects (with or without interactions) have been used; therefore, consistent measures of fit, independent of number of predictors are needed. For instance for two predictors of $(x_1, x_2)$ which are used in simple linear regression, there will be five predictors $(x_1, x_2, x_1x_2, x_1^2, x_2^2)$ for the polynomial regression. Therefore, performance evaluation of the models needs to be consistent among all models and independent of number of predictors.

RSquare ($R^2$) is a popular measurement of fit which can be calculated based on the predicted and actual response variables. The following is the formula for calculating $R^2$:

$$R^2 = 1 - \frac{\sum_{i=1}^{n}(y_i - \hat{y}_i)^2}{\sum_{i=1}^{n}(y_i - \bar{y})^2} \quad (5)$$

where $\hat{y}_i$ is the predicted value for actual $y_i$ data reading, and $\bar{y}$ is the mean of response $y_i$ readings. Root Average Square Error (RASE) is the root of average value of squared error:

$$RASE = \sqrt{\frac{\sum_{i=1}^{n}\varepsilon_i^2}{n}} = \sqrt{\frac{\sum_{i=1}^{n}(y_i - \hat{y}_i)^2}{n}}. \quad (6)$$

Average Absolute Error (AAE) can be defined as:

$$AAE = \frac{\sum_{i=1}^{n}|(y_i - \hat{y}_i)|}{n}. \quad (7)$$

For the purpose of measurement of fit and comparing model's performance, all there measures of $R^2$, RASE and AAE (formulas 5, 6 and 7) have been used in this report.

## 7   Production Simulation and Analysis

In performance analysis of the regression models, oil and natural gas exhibit different prediction success rates and behavior. In this study, the Poisson quadratic polynomial model proved to be the most realistic and accurate method for modelling oil production. Poisson distribution is used for count and rate response variables, which could be also used for oil, such as barrels per day. For predicting natural gas production application of quadratic polynomials proved to be the most accurate method, while using Poisson reduced the predictability rate slightly (Figures 3 and 4).

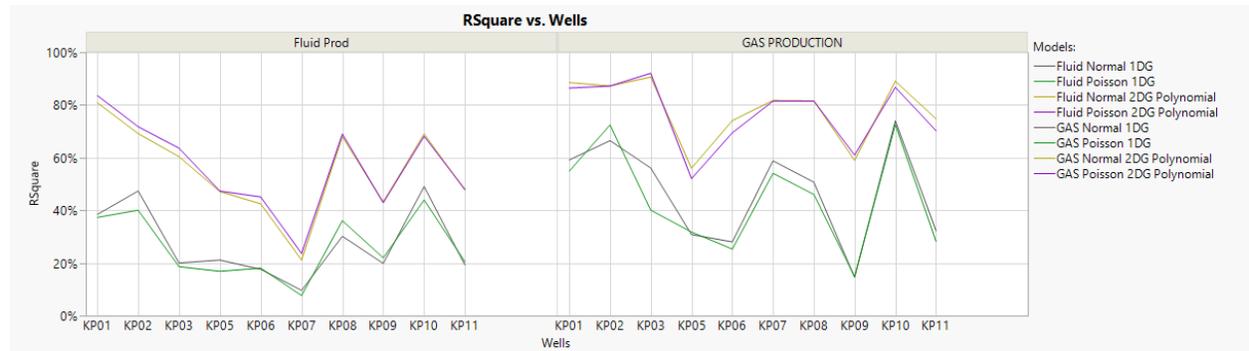

Figure 3. Comparisons of R2 for regression models used to predict fluids and natural gas productions for each well

As can be seen in Figure 4, for well KP01 all measurement of fit indicators ($R^2$, RASE and AAE) are in harmony and prove validity of the comparison benchmarks. For instance the lowest $R^2$ also shows lowest RASE, and AAE as well. Appendix A contains the model comparison tables for all 10 wells.

| Model Comparison ("KA01/KP01") | | | | | | | | | | | |
|---|---|---|---|---|---|---|---|---|---|---|---|
| **Measures of Fit for Fluid Prod** | | | | | | **Measures of Fit for GAS PRODUCTION** | | | | | |
| Predictor | .2 .4 .6 .8 | RSquare | RASE | AAE | Freq | Predictor | .2 .4 .6 .8 | RSquare | RASE | AAE | Freq |
| Fluid GLM OLS 1DG | | 0.3843 | 9.5844 | 7.3289 | 1300 | GAS GLM OLS 1DG | | 0.5913 | 5941.5 | 4398.0 | 1209 |
| Fluid GLM OLS 2DG Poly | | 0.8088 | 5.3412 | 3.9653 | 1300 | GAS GLM OLS 2DG Polynomial | | 0.8850 | 3151.2 | 2133.1 | 1209 |
| Fluid GLM Poisson 1DG | | 0.3732 | 9.6708 | 7.3647 | 1300 | GAS GLM Poisson 1DG | | 0.5491 | 6240.6 | 4641.8 | 1209 |
| Fluid GLM Poisson 2DG Poly | | 0.8363 | 4.9426 | 3.6794 | 1300 | GAS GLM Poisson 2DG Polynomial | | 0.8645 | 3421.4 | 2376.4 | 1209 |

Figure 4. Comparisons of $R^2$, RASE and AAE for regression models used to predict fluids and natural gas productions for KP01

## 7.1 Fluid Production Analysis

Figure 5 illustrates the simulation of fluid (oil and water together) production using linear regression models, first and second degree quadratic polynomials, in addition to Poisson first and second degree quadratic polynomials. Application of second degree quadratic polynomials present the most significant improvement. However, reviewing the production plot reveals a significant improvement that makes the Poisson distribution more realistic. The models with Normal distribution under certain conditions predict negative productions that is physically meaningless. These specific conditions may arise when reservoir temperatures were dropping, consequently linear systems with Normal distributions were showing downwards trends into negative territory. However, Poisson distribution, does not predict any negative production values, showed production drop towards zero line (Figure 6).

Another observation for fluid production modelling, is using the first degree regression models (both Normal and Poisson), there are the fluctuations in the actual production that models do not reproduce and may look similar to noise. However, by introducing second degree quadratic regression models, large portions of the production fluctuations were modeled. Meaning the production fluctuations were due to temperature change in the reservoir and regression models can predict the production based on reservoir temperatures more accurately, when incorporating interactions between thermocouples' observations. Appendix F contains the production prediction comparisons for all 11 wells.

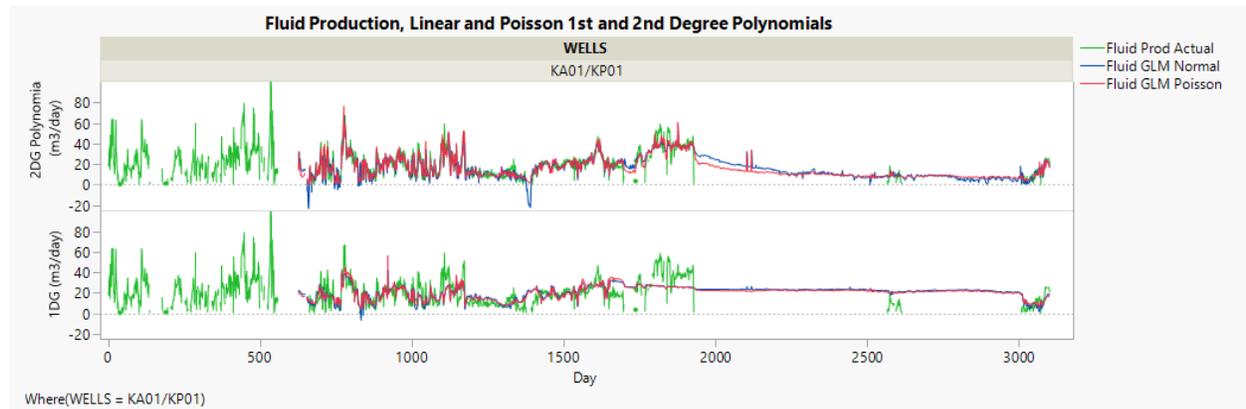

Figure 5. Comparison of Fluid production predictions versus actual for the well life cycle. Red lines indicating regression using Poisson and blue lines indicate using Normal distribution.

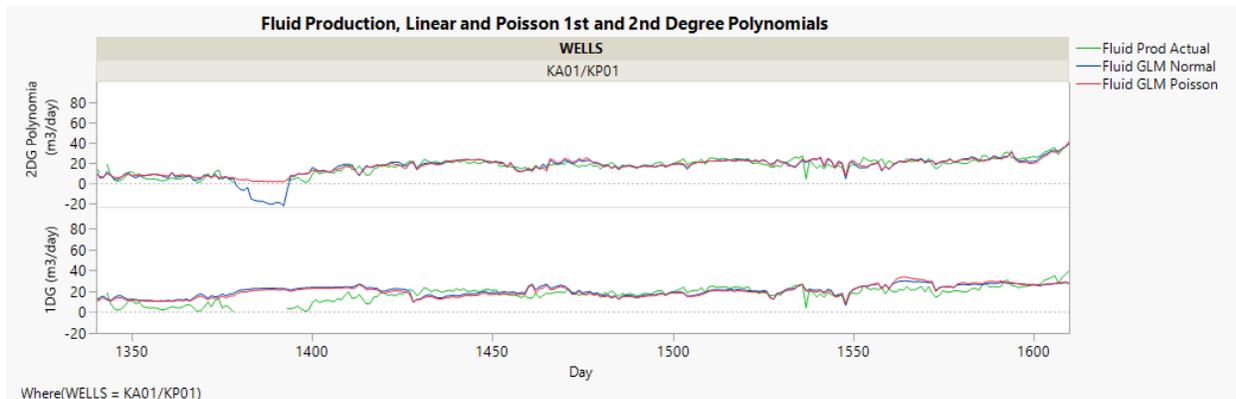

Figure 6. Comparison of Fluid production predictions versus actual for a period of 250 days. Second degree quadratic Poisson distribution creates most realistic predictions.

Table 1, illustrates the top 20 main contributing factors to build the quadratic Poisson regression model for fluid production. As can be seen the most significant predictor is thermocouple 1, which is located at the toe and can be considered Toe Temperature. Both thermocouples 10 and 11 are located at the middle and can be considered as a strong correlation with mid-temperature.

One of the main observations from Table 1 is the importance of interactions to predict fluid production. For instance interaction $(x_j x_k)$ between thermocouples 8 and 12 (or 1 and 9) have a high correlation with fluid production. The squared value $(x_j^2)$ of thermocouple 6 (similarly 7 or 9) also relatively have a high correlation, meaning there might be relationships between squared value of thermocouples and fluid production to be quantitatively explored. Appendix H contains the complete list of effects for well KP01.

| Source | LogWorth |
| --- | --- |
| THERMOCOUPLE 1 | 29.311 |
| THERMOCOUPLE 11 | 18.124 |
| THERMOCOUPLE 10 | 16.140 |
| THERMOCOUPLE 8*THERMOCOUPLE 12 | 15.856 |
| THERMOCOUPLE 1*THERMOCOUPLE 9 | 15.256 |
| THERMOCOUPLE 1*THERMOCOUPLE 8 | 14.423 |
| THERMOCOUPLE 6*THERMOCOUPLE 6 | 10.691 |
| THERMOCOUPLE 7*THERMOCOUPLE 7 | 10.138 |
| THERMOCOUPLE 12*THERMOCOUPLE 18 | 8.276 |
| THERMOCOUPLE 13 | 7.543 |
| THERMOCOUPLE 8*THERMOCOUPLE 9 | 6.984 |
| THERMOCOUPLE 9*THERMOCOUPLE 11 | 6.906 |
| THERMOCOUPLE 12*THERMOCOUPLE 17 | 6.798 |
| THERMOCOUPLE 1*THERMOCOUPLE 16 | 6.701 |
| THERMOCOUPLE 9*THERMOCOUPLE 9 | 6.593 |
| THERMOCOUPLE 3*THERMOCOUPLE 12 | 6.573 |
| THERMOCOUPLE 1*THERMOCOUPLE 20 | 6.431 |
| THERMOCOUPLE 4*THERMOCOUPLE 18 | 6.376 |
| THERMOCOUPLE 3*THERMOCOUPLE 11 | 6.150 |
| THERMOCOUPLE 3*THERMOCOUPLE 17 | 5.812 |

Table 1. Effect Summary, top 20 most effective factors where, $LogWorth = -log_{10}(PValue)$

## 7.2  Natural Gas Production Analysis

Figure 7 illustrates the overall natural gas predictions by the models compared with the actual production. This comparison contains the first degree regressions using Poisson and Normal distributions, along with second degree quadratic polynomials. Appendix G contains the production predictions for all 10 well.

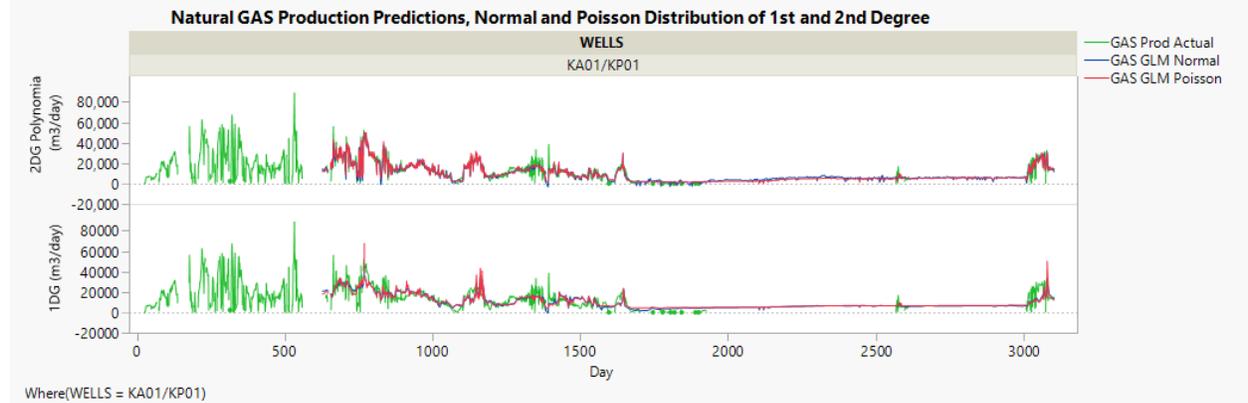

Figure 7. Comparisons of Natural Gas production predictions versus actual.

Figure 8 illustrates the natural gas productions from day 1350 to 1600 for more detailed analysis. As can be seen second degree quadratic polynomial with Poisson distribution offers a more realistic prediction model, by restricting the prediction to non-negative values, which is a major characteristic of Poisson distribution.

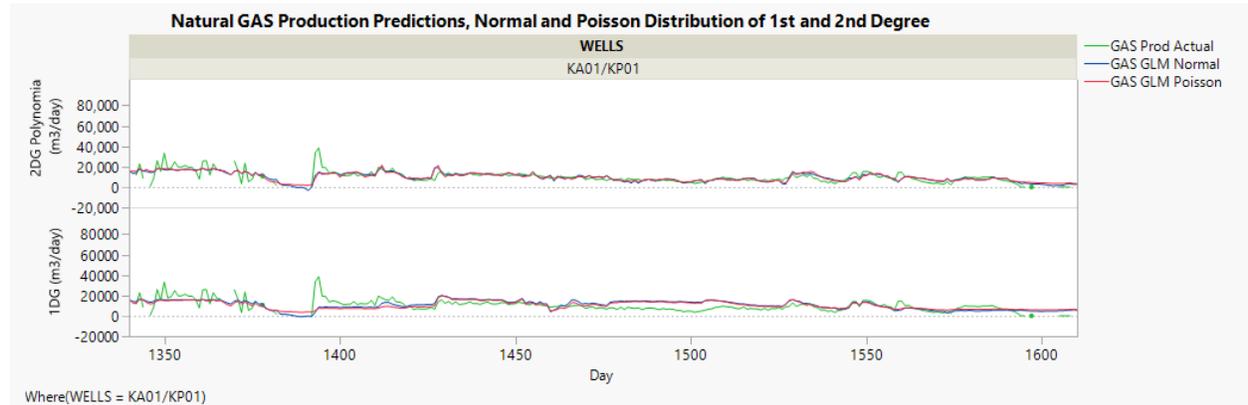

Figure 8. Comparisons of Natural Gas production predictions versus actual for KP01 from day 1350 to 1600.

Table 2 illustrates the top 20 most contributing factors for building natural gas predictive model with quadratic Poisson regression. It worth noting stronger presence of interactions compare to the similar model for predicting fluids. Also the most effective thermocouples for predicting the natural gas are closer to mid or heel sections. Appendix I contains the complete list of effects analysis for KP01.

| Source | LogWorth |
| --- | --- |
| THERMOCOUPLE 10*THERMOCOUPLE 12 | 14192.84 |
| THERMOCOUPLE 10*THERMOCOUPLE 13 | 12668.21 |
| THERMOCOUPLE 11*THERMOCOUPLE 12 | 9948.348 |
| THERMOCOUPLE 14 | 9040.609 |
| THERMOCOUPLE 10*THERMOCOUPLE 14 | 7235.084 |
| THERMOCOUPLE 13 | 6822.531 |
| THERMOCOUPLE 12*THERMOCOUPLE 20 | 6646.944 |

| Source | LogWorth |
|---|---|
| THERMOCOUPLE 3 | 6041.172 |
| THERMOCOUPLE 8*THERMOCOUPLE 13 | 5462.259 |
| THERMOCOUPLE 11*THERMOCOUPLE 20 | 5151.365 |
| THERMOCOUPLE 10*THERMOCOUPLE 16 | 5043.121 |
| THERMOCOUPLE 8*THERMOCOUPLE 17 | 4964.610 |
| THERMOCOUPLE 3*THERMOCOUPLE 11 | 4876.550 |
| THERMOCOUPLE 11*THERMOCOUPLE 16 | 4775.787 |
| THERMOCOUPLE 11*THERMOCOUPLE 14 | 4678.654 |
| THERMOCOUPLE 11*THERMOCOUPLE 13 | 4090.118 |
| THERMOCOUPLE 12*THERMOCOUPLE 18 | 3954.697 |
| THERMOCOUPLE 17 | 3549.411 |
| THERMOCOUPLE 8*THERMOCOUPLE 18 | 3400.160 |
| THERMOCOUPLE 12 | 3383.417 |

Table 2. Effect Summary, top 20 most effective factors for second degree quadratic polynomial with Poisson distribution ($LogWorth = -log_{10}(PValue)$)

## 8   Analysis of Residuals

The residual plot is a method of validating and observing the strength of predictive models. The residuals defined as the difference between observed and predicted values:

$$\varepsilon_i = y_i - \hat{y}_i$$

where $\varepsilon_i$ is the residual, $y_i$ is actual observation and $\hat{y}_i$ is the predicted or fitted value. Ideally the residuals should consist only of random unpredictability, independent from response value, $y_i$ or $\hat{y}_i$ and predictors such as $x_i$. Therefore, distribution probability function of residuals should form a Normal distribution. Any deviation from Normal distribution, indicate presence of bias within the regression model such as missing some predictors, interactions or higher degrees of polynomials may be required for the model.

Figure 10 illustrates the scatter plot of predicted versus actual. In ideal conditions, the predicted versus actual should form a straight $45°$ diagonal line (as illustrated by blue line). However, due to existence of noise, measurement errors, and other unknown reasons, the scatter plot may contain large variations from the diagonal line. Figure 10 demonstrates predictability performance improvements by using second degree quadratic polynomials using both Normal and Poisson distributions. Numerical comparisons of measurements of fit displayed in Figure 4, demonstrate clear prediction improvements by using quadratic polynomials and slight improvements due to application of Poisson distribution. Appendix B contains the comparison plots of actuals versus predicted for all 10 wells.

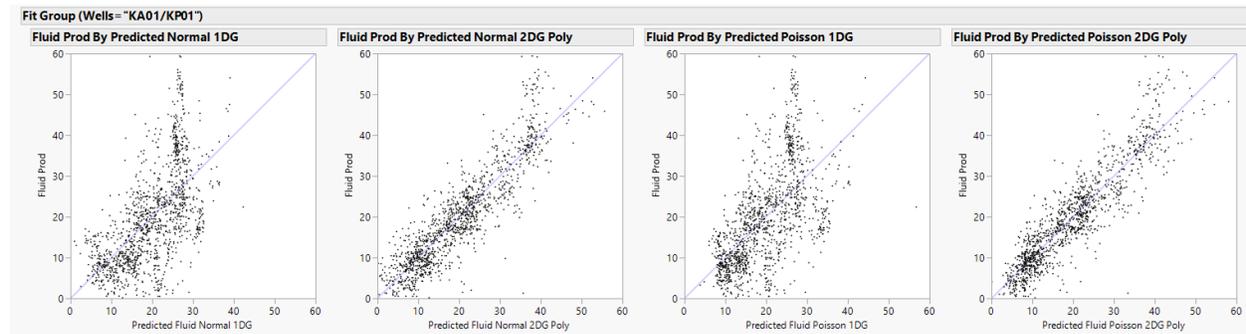

Figure 10. Bivariate plot of fluid production, predicted versus actual

Figure 11 is scatter plot of predicted versus actual for natural gas production regression models. Similar to predictions of fluids, the natural gas predictions exhibit major improvement by using second degree quadratic regression polynomials, while, application of Poisson distribution decreased the predictability marginally. Appendix C contains the comparison plots of actuals versus predicted for all 10 wells.

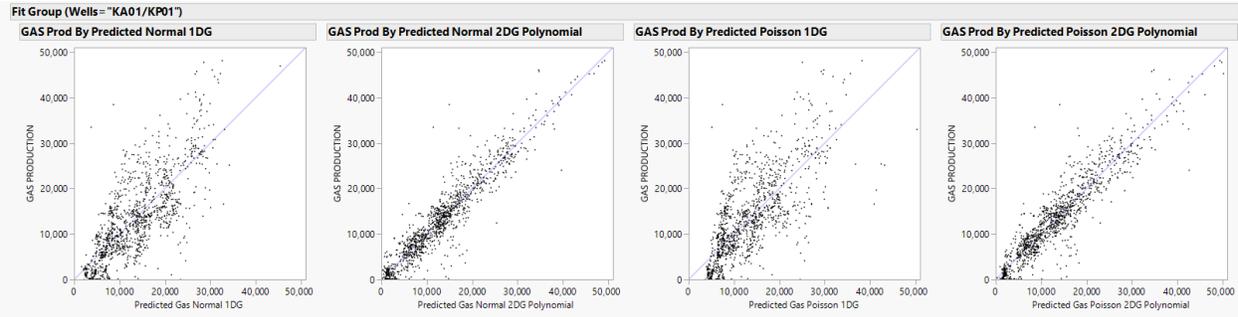

Figure 11. Bivariate plot of natural gas production, predicted versus actual

### 8.1 Analysis of Fluid Production Residual Histograms

Figure 12 illustrates the residual distribution of fluid production predictions for well KP01. Ideally the residuals should form a Normal distribution as shown by red line. However, practically the residuals may not fully conform to the Normal distribution. As can be seen in figure 12, the actual residuals may deviate slightly from the Normal distributions; however, as illustrated the residuals follow Normal distribution very closely and there are not any major skewness. The topic of quantifying and numerically analyzing residual skewness remains to be addressed in future investigations. Appendix D contains residual histograms for all 10 wells.

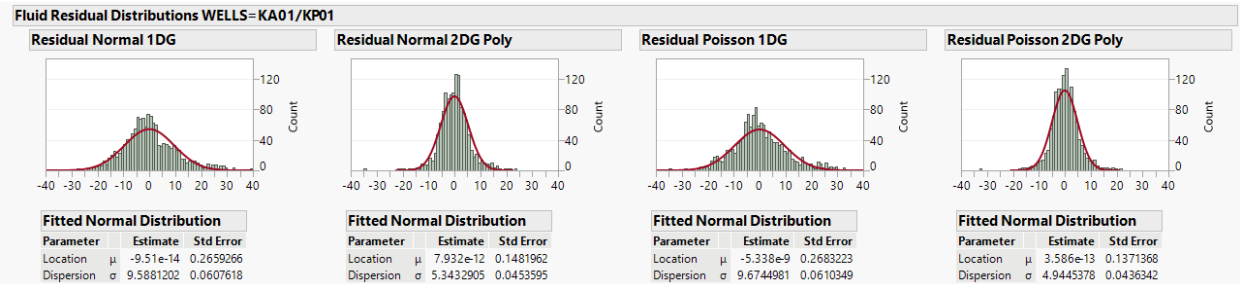

Figure 12. Histogram of fluid production prediction residuals vs Normal distribution (red line)

### 8.2 7.2. Analysis of Gas Production Residual Histograms

Figure 13 illustrates the distribution analysis of natural gas production residuals for KP01. This pattern is very consistent with other 9 wells as well, which are included in Appendix E. The average ($\mu$) of residuals are statistically very close to zero, meaning the regression models are coherent and have been calculated correctly. However, dispersions, or standard error of average ($\mu$) increased by introduction of Poisson distribution, meaning the models' accuracy have decreased. The best predictions for natural gas production were obtained by application of second degree quadratic polynomials.

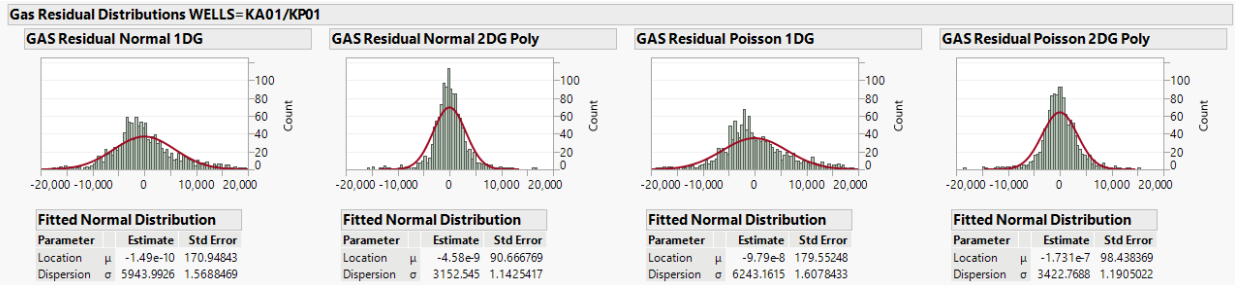
Figure 13. Histogram of Gas production prediction residuals versus Normal distribution (red line)

## 9 Conclusion

The most modelling performance improvements were achieved by adding two-way interactions between thermocouples for both fluids and natural gas production models. Temperature interactions play a major role in predicting the production, perhaps might be related to presence of reservoir temperature increase.

Adding Poisson distribution improved the overall oil prediction performance slightly; however, eliminated a major issue with predicting negative values which are created by Normal distribution linear regression. The main reason for resolving negative prediction values are related to the nature of Poisson distribution for predicting non-negative event, or rates.

Residual histograms for both fluid and natural gas are very close to Normal distribution, meaning regression models are coherent. Therefore, showing THAI process can be modelled by only using temperatures with a relatively high degree of success, and the nature of process is second degree polynomial.

Analysis of outliers is a critical phases of data analytics initiative. Prior to analysis all temperature readings above 700 °C were reduced to be within physical range of THAI operations. A comprehensive multivariable outlier analysis such as Mahalanobis was performed and resulted in removal of a small number of observations. However, more research in systematic analysis of multivariable outliers are needed.

## 10 Future Research Opportunities

Elements of time were not included in building the predictive models in this paper. Time will have deep effects on heat propagation affecting the reservoir behavior. Time series analysis including covariances such as air injection volumes, Bottom Hole Pressure, and chemical composition and their interactions are needed for more detailed analysis of THAI process.

Spatial variations in fluid properties such as oil viscosity, deeply affects the models and predictability success and need to be included in future analysis. For instance throughout a reservoir, oil viscosity naturally may have variations which should be considered as a predictors within the models.

Application of Poisson distribution has the restriction of mean and variance to be equal. Most practical applications offer situations that variance is larger than mean (over-dispersion). Effects of over-dispersion need to be analysed and included within the reservoir models.

In a machine learning project, the data is usually divided into training and validation segments. Training segment is used for building the models and validation for testing the models on data that was not used for training. In this report due to number of models being compared all data was used for training and the overall model performance was considered. More in-depth analysis with a smaller number of models is required to validate the performance measures using the data that was not used for training.

Poisson is a special case of larger family of distributions such as exponential and gamma probability density functions. More experiments with other probability density functions, using larger datasets are required to explore the most effective distributions for modelling THAI production process.

Residual analysis presented in figures 12 and 13 exhibit some small level of skewness compare to Normal distribution. The skewness might contain some information to improve the models predicting the thermal process more accurately. More in-depth analysis of the residuals may improve the understanding and exact nature of THAI process in more details.

## 11 Acknowledgements

## 12 References

[1] Priyanka Sinha, Multivariate Polynomial Regression in Data Mining: Methodology, Problems and Solutions, International Journal of Scientific & Engineering Research, Volume 4, Issue 12, December-2013, ISSN 2229-551

[2] W. Wei, J. Wang and I. Gates, New Insights into Kerrobert Toe-to-Heel Air Injection (THAI) Production Analysis, 81st EAGE Conference and Exhibition 2019 Conference Proceedings, Jun 2019, Volume 2019, p.1 - 5

[3] Muhammad Rabiu Ado, Malcolm Greaves, Sean P. Rigby, Effect of pre-ignition heating cycle method, air injection flux, and reservoir viscosity on the THAI heavy oil recovery process, Journal of Petroleum Science and Engineering, ELSEVIER, Volume 166, July 2018, p. 94-103

[4] S. Larter, J. Adams, I.D. Gates, B. Bennett, H. Huang, The Origin, Prediction and Impact of Oil Viscosity Heterogeneity on the Production Characteristics of Tar Sand and Heavy Oil Reservoirs, Journal of Canadian Petroleum Technology, Volume 47, Issue 01, January 2008, p. 52-61

[5] JMP, A Business Unit of SAS, JMP® 15 Fitting Linear Models, SAS Institute, Cary, NC, USA, 2020, p. 47

[6] Shalabh, MTH 416: Regression Analysis Course Material, Chapter 15: Poisson Regression Models, Indian Institute of Technology Kanpur, India, Department of Mathematics & Statistics, National Digital Library of India, 2020

[7] L. Simon, R. Heckard, A. Wiesner and D. Young, STAT501: Regression Methods, Online Notes, The Pennsylvania State University, Eberly College of Science, Chapter 15 Logistics, Poisson and Nonlinear Regression, 2020

[8] Jon Wakefield, Bayesian and Frequentist Regression Methods, Springer, New York, 2013, ISBN 978-1-4419-0925-1, Chapter 6: General Regression Models

[9] J. Rosenberger, STAT 503: Design of Experiments, Online Notes, The Pennsylvania State University, Eberly College of Science, Lesson 11: Response Surface Methods and Designs, 2020

[10] C. Robinson and R. E. Schumacker, Interaction Effects: Centering, Variance Inflation Factor, and Interpretation Issues, Multiple Linear Regression Viewpoints, 2009, 35(1):6-11

[11] SAS Institute Inc., Usage Note 60335: Choice of continuous response distribution in log-linked GLMs, SAS Technical Support, 2017

**Appendix A – Model Comparisons**

Comparison table for the models used in the study. Ordinary Least Square (OLS) uses Normal distribution.

**Model Comparison ("KA01/KP01")**

Measures of Fit for Fluid Prod

| Predictor | .2 .4 .6 .8 | RSquare | RASE | AAE | Freq |
|---|---|---|---|---|---|
| Fluid GLM OLS 1DG | | 0.3848 | 9.5839 | 7.3219 | 1295 |
| Fluid GLM OLS 2DG Poly | | 0.8084 | 5.3487 | 3.9712 | 1295 |
| Fluid GLM Poisson 1DG | | 0.3739 | 9.6681 | 7.3544 | 1295 |
| Fluid GLM Poisson 2DG Poly | | 0.8359 | 4.9500 | 3.6862 | 1295 |

Measures of Fit for GAS PRODUCTION

| Predictor | .2 .4 .6 .8 | RSquare | RASE | AAE | Freq |
|---|---|---|---|---|---|
| GAS GLM OLS 1DG | | 0.5902 | 5927.3 | 4393.4 | 1204 |
| GAS GLM OLS 2DG Polynomial | | 0.8840 | 3153.6 | 2134.0 | 1204 |
| GAS GLM Poisson 1DG | | 0.5466 | 6235.0 | 4640.0 | 1204 |
| GAS GLM Poisson 2DG Polynomial | | 0.8632 | 3425.2 | 2379.7 | 1204 |

**Model Comparison ("KA02/KP02")**

Measures of Fit for Fluid Prod

| Predictor | .2 .4 .6 .8 | RSquare | RASE | AAE | Freq |
|---|---|---|---|---|---|
| Fluid GLM OLS 1DG | | 0.4735 | 5.9334 | 3.8792 | 996 |
| Fluid GLM OLS 2DG Poly | | 0.6917 | 4.5402 | 2.9242 | 996 |
| Fluid GLM Poisson 1DG | | 0.4004 | 6.3319 | 4.1933 | 996 |
| Fluid GLM Poisson 2DG Poly | | 0.7177 | 4.3448 | 2.8735 | 996 |

Measures of Fit for GAS PRODUCTION

| Predictor | .2 .4 .6 .8 | RSquare | RASE | AAE | Freq |
|---|---|---|---|---|---|
| GAS GLM OLS 1DG | | 0.6749 | 8788.2 | 5852.5 | 1013 |
| GAS GLM OLS 2DG Polynomial | | 0.8721 | 5512.3 | 3938.7 | 1013 |
| GAS GLM Poisson 1DG | | 0.7449 | 7786.1 | 5230.5 | 1013 |
| GAS GLM Poisson 2DG Polynomial | | 0.8716 | 5523.4 | 3937.3 | 1013 |

**Model Comparison ("KA03/KP03")**

Measures of Fit for Fluid Prod

| Predictor | .2 .4 .6 .8 | RSquare | RASE | AAE | Freq |
|---|---|---|---|---|---|
| Fluid GLM OLS 1DG | | 0.2003 | 6.7145 | 5.0663 | 1334 |
| Fluid GLM OLS 2DG Poly | | 0.6039 | 4.7255 | 3.5350 | 1334 |
| Fluid GLM Poisson 1DG | | 0.1865 | 6.7721 | 5.1281 | 1334 |
| Fluid GLM Poisson 2DG Poly | | 0.6362 | 4.5285 | 3.3311 | 1334 |

Measures of Fit for GAS PRODUCTION

| Predictor | .2 .4 .6 .8 | RSquare | RASE | AAE | Freq |
|---|---|---|---|---|---|
| GAS GLM OLS 1DG | | 0.5615 | 11790 | 6626.2 | 1418 |
| GAS GLM OLS 2DG Polynomial | | 0.9057 | 5466.5 | 3096.7 | 1418 |
| GAS GLM Poisson 1DG | | 0.4013 | 13777 | 6884.7 | 1418 |
| GAS GLM Poisson 2DG Polynomial | | 0.9203 | 5027.5 | 2803.9 | 1418 |

**Model Comparison ("KA05/KP05")**

Measures of Fit for Fluid Prod

| Predictor | .2 .4 .6 .8 | RSquare | RASE | AAE | Freq |
|---|---|---|---|---|---|
| Fluid GLM OLS 1DG | | 0.2114 | 12.037 | 8.5501 | 1412 |
| Fluid GLM OLS 2DG Poly | | 0.4714 | 9.8550 | 6.6791 | 1412 |
| Fluid GLM Poisson 1DG | | 0.1686 | 12.359 | 8.6402 | 1412 |
| Fluid GLM Poisson 2DG Poly | | 0.4733 | 9.8371 | 6.6319 | 1412 |

Measures of Fit for GAS PRODUCTION

| Predictor | .2 .4 .6 .8 | RSquare | RASE | AAE | Freq |
|---|---|---|---|---|---|
| GAS GLM OLS 1DG | | 0.3083 | 4714.7 | 3763.9 | 1403 |
| GAS GLM OLS 2DG Polynomial | | 0.5601 | 3759.9 | 2808.4 | 1403 |
| GAS GLM Poisson 1DG | | 0.3177 | 4682.5 | 3748.4 | 1403 |
| GAS GLM Poisson 2DG Polynomial | | 0.5205 | 3925.6 | 2945.0 | 1403 |

**Model Comparison ("KA06/KP06")**

Measures of Fit for Fluid Prod

| Predictor | .2 .4 .6 .8 | RSquare | RASE | AAE | Freq |
|---|---|---|---|---|---|
| Fluid GLM OLS 1DG | | 0.1765 | 9.5621 | 6.8522 | 1120 |
| Fluid GLM OLS 2DG Poly | | 0.4248 | 7.9911 | 5.6908 | 1120 |
| Fluid GLM Poisson 1DG | | 0.1808 | 9.5369 | 6.8525 | 1120 |
| Fluid GLM Poisson 2DG Poly | | 0.4510 | 7.8072 | 5.5577 | 1120 |

Measures of Fit for GAS PRODUCTION

| Predictor | .2 .4 .6 .8 | RSquare | RASE | AAE | Freq |
|---|---|---|---|---|---|
| GAS GLM OLS 1DG | | 0.2951 | 10465 | 7602.5 | 1068 |
| GAS GLM OLS 2DG Polynomial | | 0.7411 | 6342.8 | 4240.2 | 1068 |
| GAS GLM Poisson 1DG | | 0.2684 | 10661 | 7708.4 | 1068 |
| GAS GLM Poisson 2DG Polynomial | | 0.6955 | 6878.1 | 4516.9 | 1068 |

**Model Comparison ("KA07/KP07")**

Measures of Fit for Fluid Prod

| Predictor | .2 .4 .6 .8 | RSquare | RASE | AAE | Freq |
|---|---|---|---|---|---|
| Fluid GLM OLS 1DG | | 0.0961 | 6.9823 | 4.6218 | 1982 |
| Fluid GLM OLS 2DG Poly | | 0.2115 | 6.5213 | 4.1670 | 1982 |
| Fluid GLM Poisson 1DG | | 0.0766 | 7.0575 | 4.6742 | 1982 |
| Fluid GLM Poisson 2DG Poly | | 0.2362 | 6.4183 | 4.1019 | 1982 |

Measures of Fit for GAS PRODUCTION

| Predictor | .2 .4 .6 .8 | RSquare | RASE | AAE | Freq |
|---|---|---|---|---|---|
| GAS GLM OLS 1DG | | 0.5881 | 5312.1 | 3347.8 | 1937 |
| GAS GLM OLS 2DG Polynomial | | 0.8174 | 3536.7 | 2472.6 | 1937 |
| GAS GLM Poisson 1DG | | 0.5407 | 5609.6 | 3372.6 | 1937 |
| GAS GLM Poisson 2DG Polynomial | | 0.8148 | 3561.7 | 2525.4 | 1937 |

## Model Comparison ("KA08/KP08")

### Measures of Fit for Fluid Prod

| Predictor | .2 .4 .6 .8 | RSquare | RASE | AAE | Freq |
|---|---|---|---|---|---|
| Fluid GLM OLS 1DG | | 0.3014 | 6.9963 | 5.5558 | 427 |
| Fluid GLM OLS 2DG Poly | | 0.6795 | 4.7384 | 3.3148 | 427 |
| Fluid GLM Poisson 1DG | | 0.3608 | 6.6920 | 5.1067 | 427 |
| Fluid GLM Poisson 2DG Poly | | 0.6897 | 4.6627 | 3.2760 | 427 |

### Measures of Fit for GAS PRODUCTION

| Predictor | .2 .4 .6 .8 | RSquare | RASE | AAE | Freq |
|---|---|---|---|---|---|
| GAS GLM OLS 1DG | | 0.5224 | 9111.5 | 6495.6 | 428 |
| GAS GLM OLS 2DG Polynomial | | 0.8157 | 5659.3 | 3756.4 | 428 |
| GAS GLM Poisson 1DG | | 0.4874 | 9439.1 | 6465.6 | 428 |
| GAS GLM Poisson 2DG Polynomial | | 0.8147 | 5675.4 | 3907.7 | 428 |

## Model Comparison ("KA09/KP09")

### Measures of Fit for Fluid Prod

| Predictor | .2 .4 .6 .8 | RSquare | RASE | AAE | Freq |
|---|---|---|---|---|---|
| Fluid GLM OLS 1DG | | 0.1986 | 5.6495 | 4.3863 | 1319 |
| Fluid GLM OLS 2DG Poly | | 0.4314 | 4.7587 | 3.5955 | 1319 |
| Fluid GLM Poisson 1DG | | 0.2198 | 5.5745 | 4.3114 | 1319 |
| Fluid GLM Poisson 2DG Poly | | 0.4292 | 4.7680 | 3.6468 | 1319 |

### Measures of Fit for GAS PRODUCTION

| Predictor | .2 .4 .6 .8 | RSquare | RASE | AAE | Freq |
|---|---|---|---|---|---|
| GAS GLM OLS 1DG | | 0.1455 | 6047.0 | 4280.5 | 1349 |
| GAS GLM OLS 2DG Polynomial | | 0.5899 | 4188.9 | 2676.2 | 1349 |
| GAS GLM Poisson 1DG | | 0.1492 | 6033.8 | 4257.9 | 1349 |
| GAS GLM Poisson 2DG Polynomial | | 0.6100 | 4085.2 | 2592.8 | 1349 |

## Model Comparison ("KA10/KP10")

### Measures of Fit for Fluid Prod

| Predictor | .2 .4 .6 .8 | RSquare | RASE | AAE | Freq |
|---|---|---|---|---|---|
| Fluid GLM OLS 1DG | | 0.4903 | 5.3660 | 3.9035 | 457 |
| Fluid GLM OLS 2DG Poly | | 0.6902 | 4.1831 | 2.9835 | 457 |
| Fluid GLM Poisson 1DG | | 0.4395 | 5.6271 | 4.1516 | 457 |
| Fluid GLM Poisson 2DG Poly | | 0.6813 | 4.2433 | 3.0146 | 457 |

### Measures of Fit for GAS PRODUCTION

| Predictor | .2 .4 .6 .8 | RSquare | RASE | AAE | Freq |
|---|---|---|---|---|---|
| GAS GLM OLS 1DG | | 0.7398 | 2724.0 | 1736.6 | 445 |
| GAS GLM OLS 2DG Polynomial | | 0.8907 | 1765.5 | 1129.4 | 445 |
| GAS GLM Poisson 1DG | | 0.7254 | 2798.1 | 1866.7 | 445 |
| GAS GLM Poisson 2DG Polynomial | | 0.8672 | 1945.7 | 1278.4 | 445 |

## Model Comparison ("KA11/KP11")

### Measures of Fit for Fluid Prod

| Predictor | .2 .4 .6 .8 | RSquare | RASE | AAE | Freq |
|---|---|---|---|---|---|
| Fluid GLM OLS 1DG | | 0.1926 | 6.8317 | 5.1253 | 943 |
| Fluid GLM OLS 2DG Poly | | 0.4765 | 5.5008 | 3.8847 | 943 |
| Fluid GLM Poisson 1DG | | 0.2046 | 6.7805 | 5.0768 | 943 |
| Fluid GLM Poisson 2DG Poly | | 0.4796 | 5.4846 | 3.8639 | 943 |

### Measures of Fit for GAS PRODUCTION

| Predictor | .2 .4 .6 .8 | RSquare | RASE | AAE | Freq |
|---|---|---|---|---|---|
| GAS GLM OLS 1DG | | 0.3227 | 7071.6 | 5012.5 | 969 |
| GAS GLM OLS 2DG Polynomial | | 0.7472 | 4320.1 | 2850.7 | 969 |
| GAS GLM Poisson 1DG | | 0.2821 | 7280.5 | 5155.5 | 969 |
| GAS GLM Poisson 2DG Polynomial | | 0.7023 | 4688.7 | 3295.2 | 969 |

## Appendix B – Fluid Production Predicted versus Actual Plot

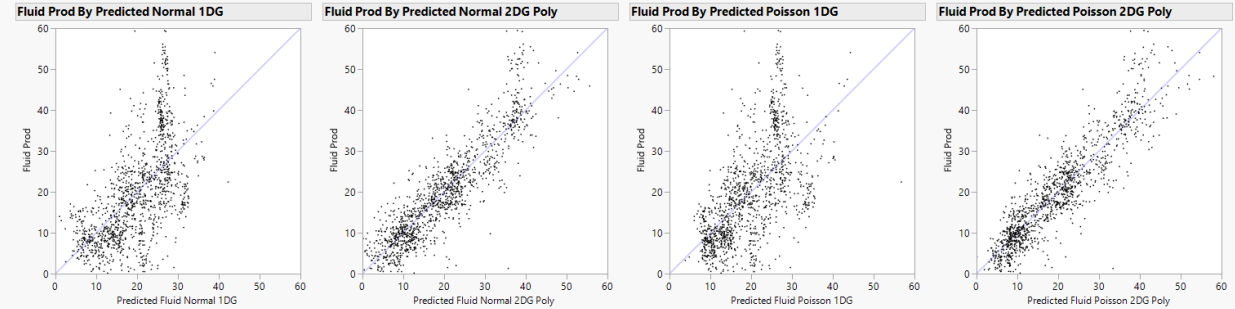

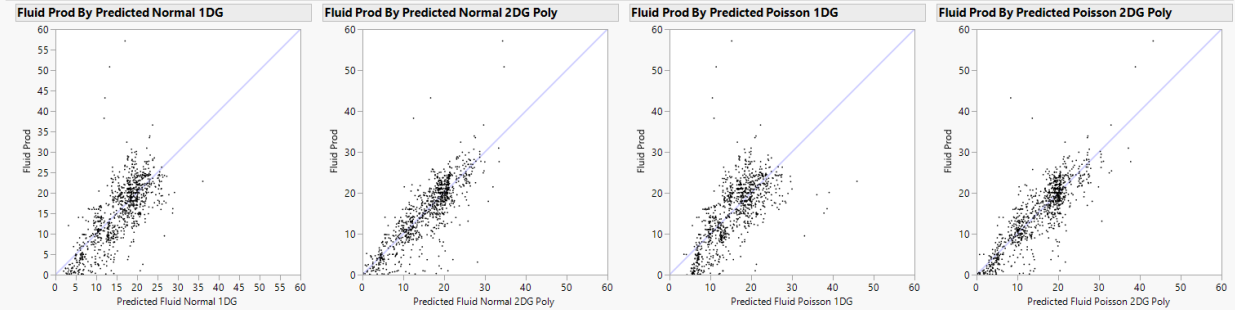

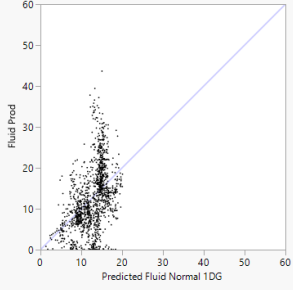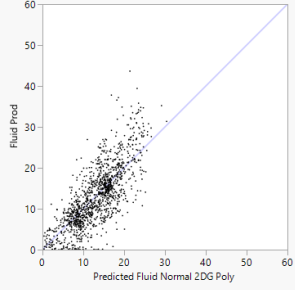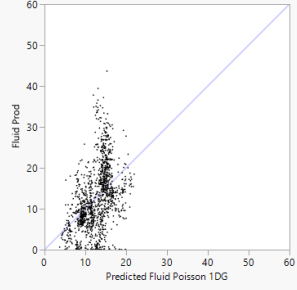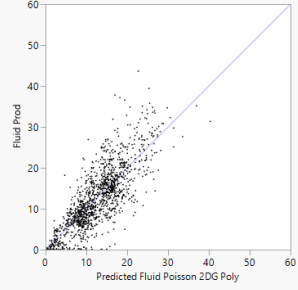
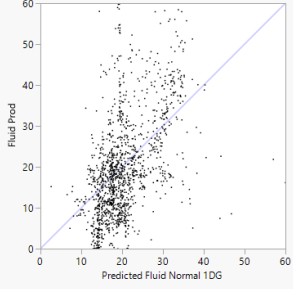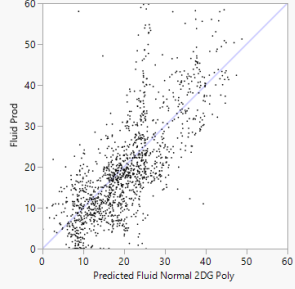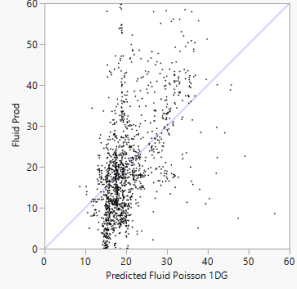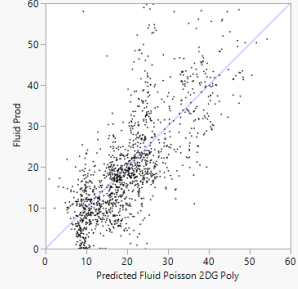
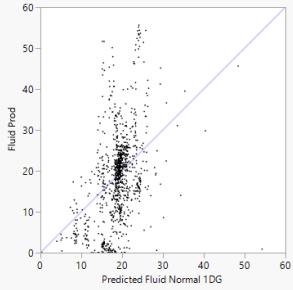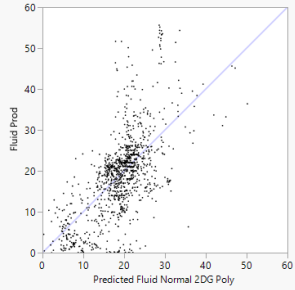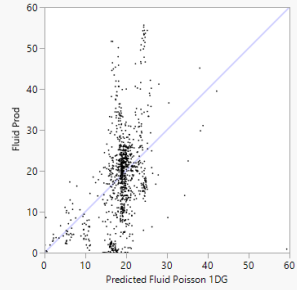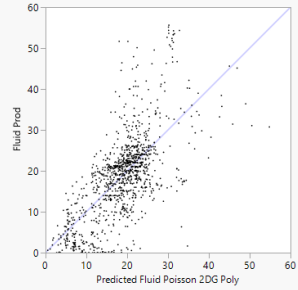
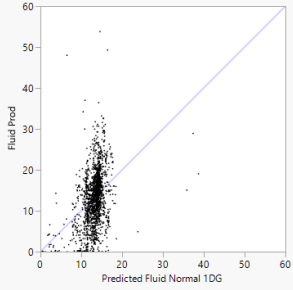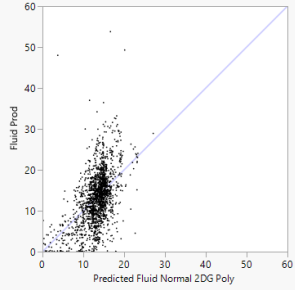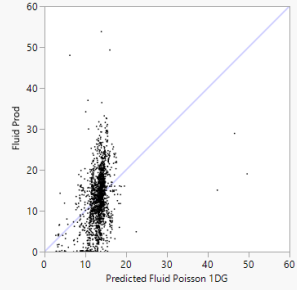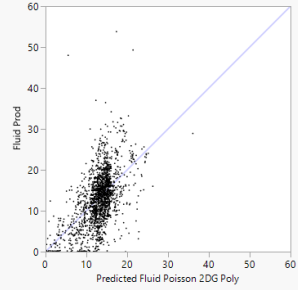

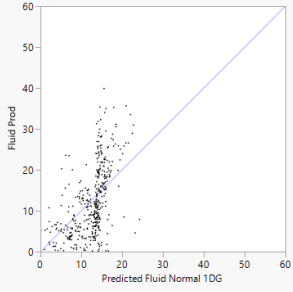 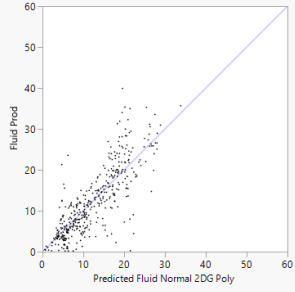 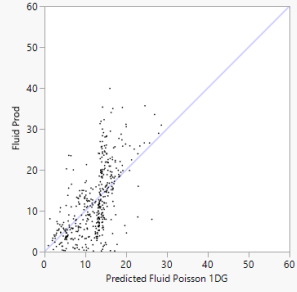 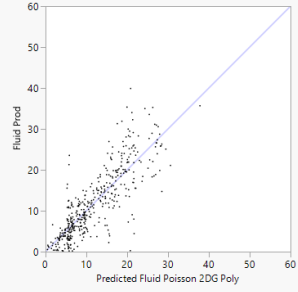
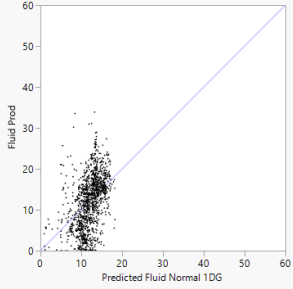 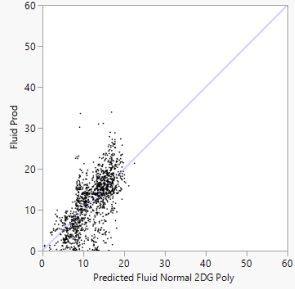 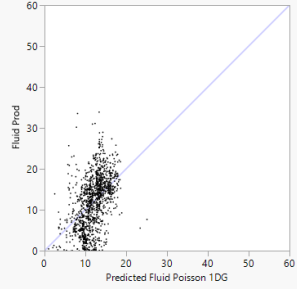 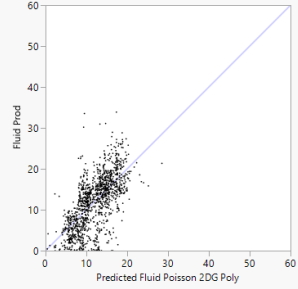
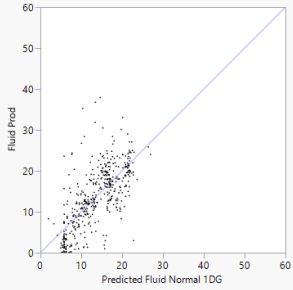 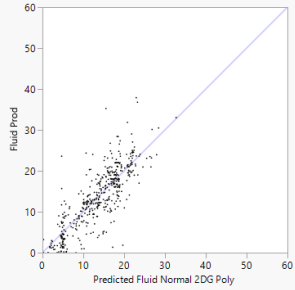 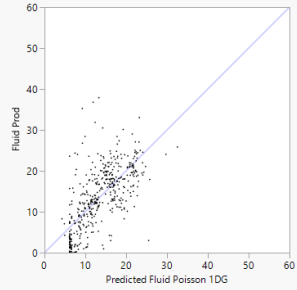 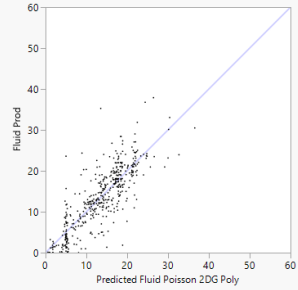
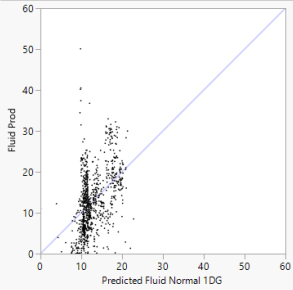 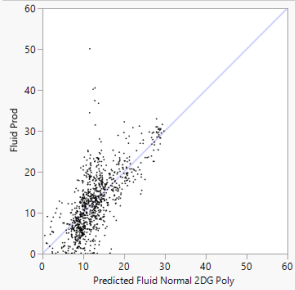 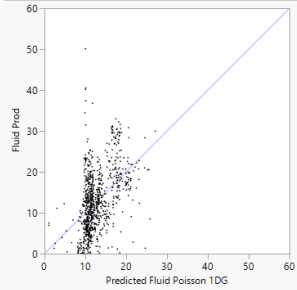 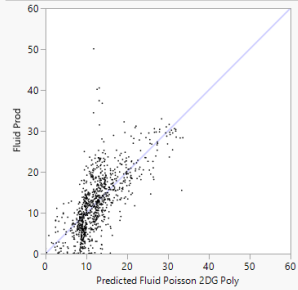

# Appendix C – Natural Gas Production Predicted versus Actual Plot

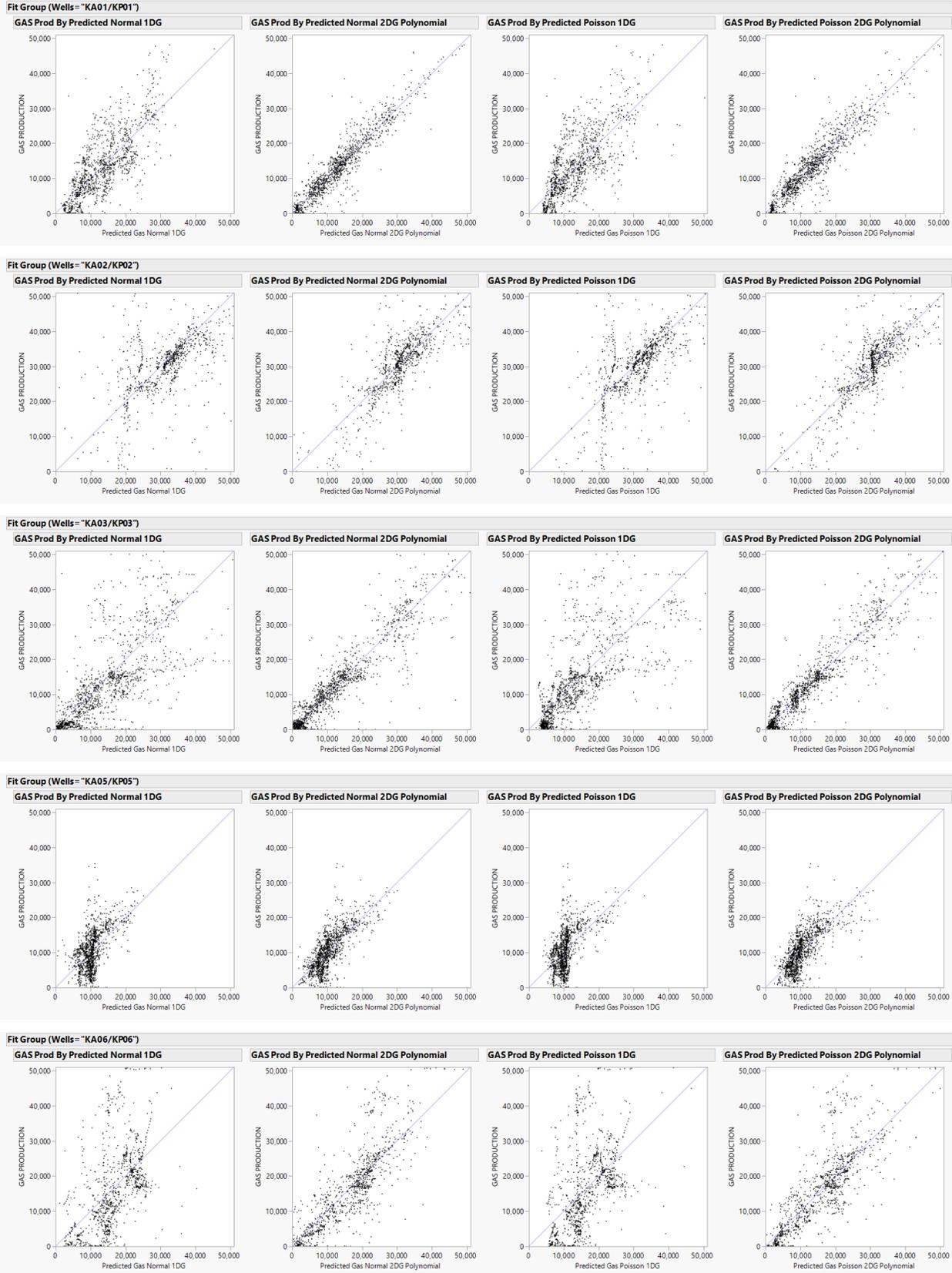

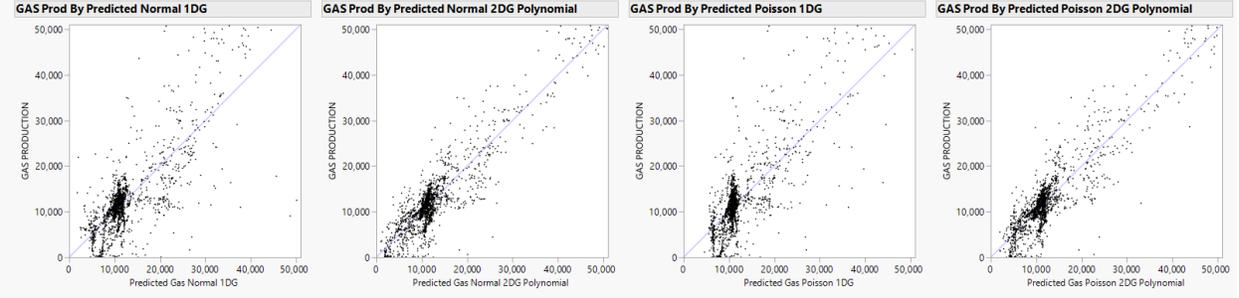
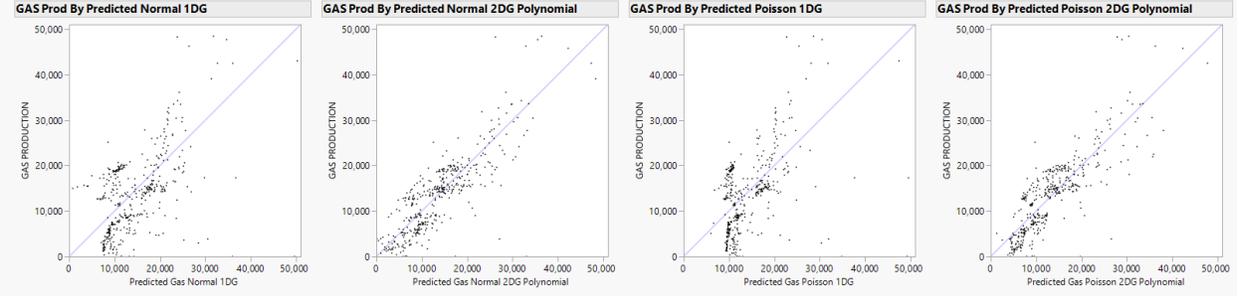
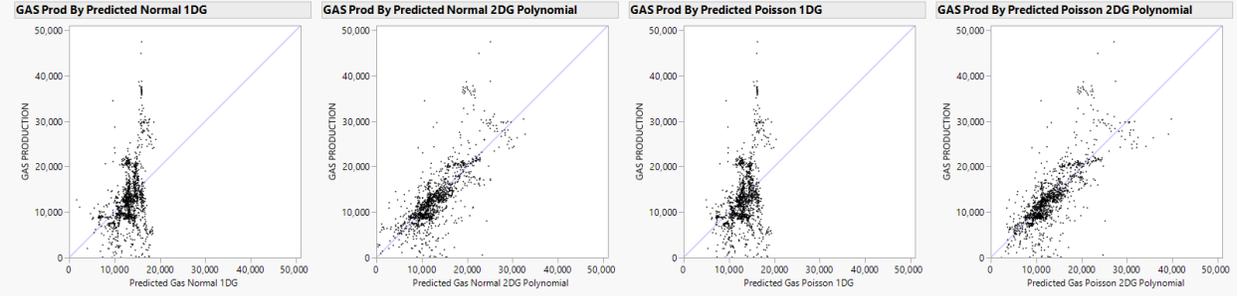
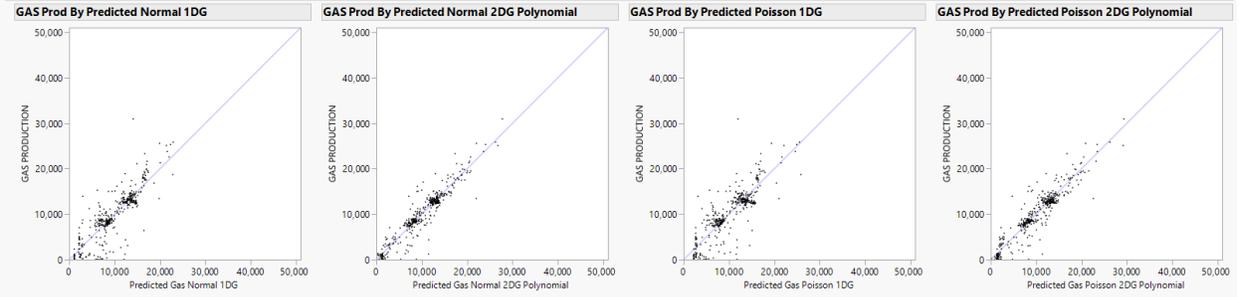
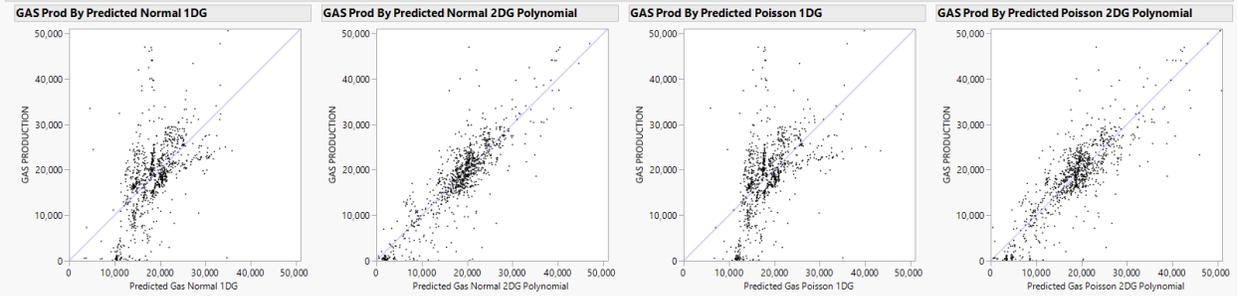

# Appendix D – Residual Distribution Analysis of Fluid Predictions

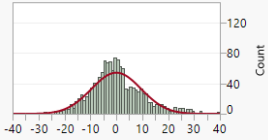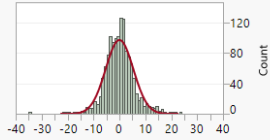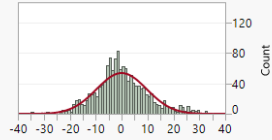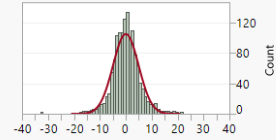
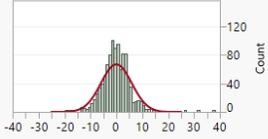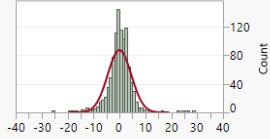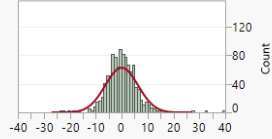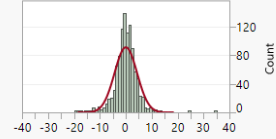
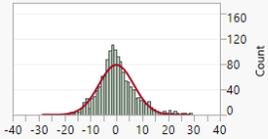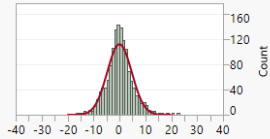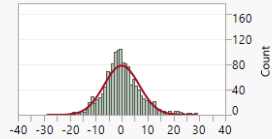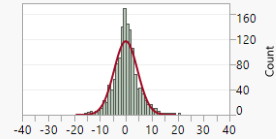
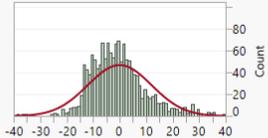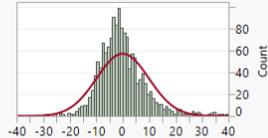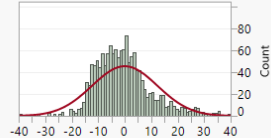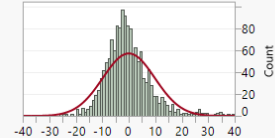
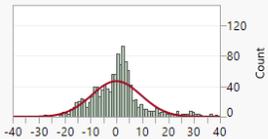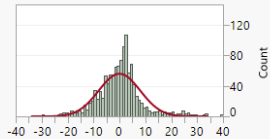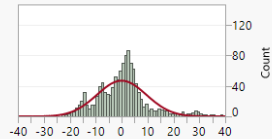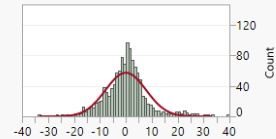

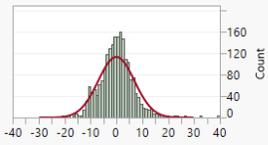 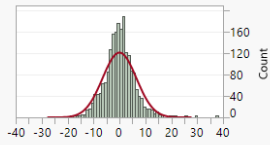 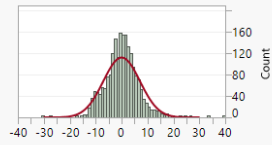 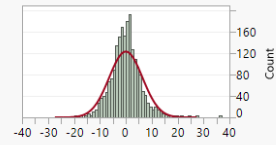
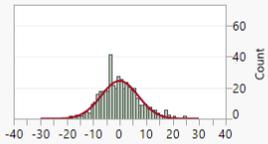 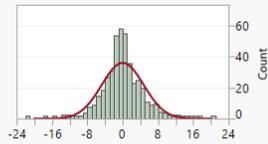 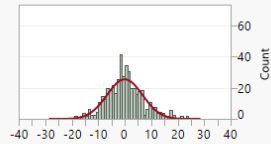 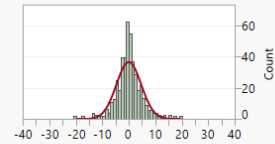
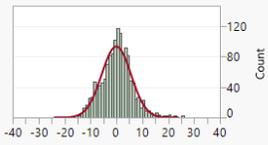 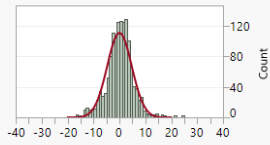 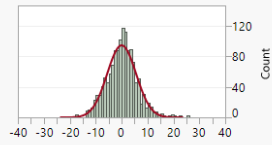 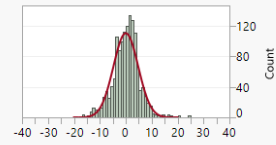
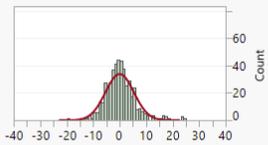 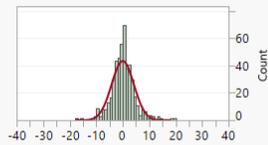 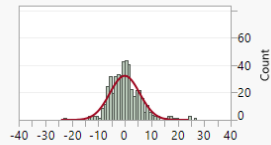 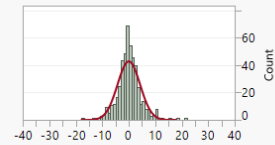
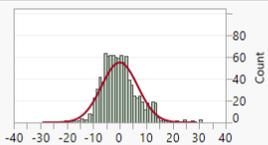 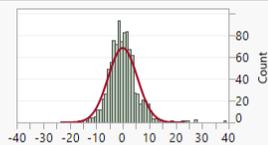 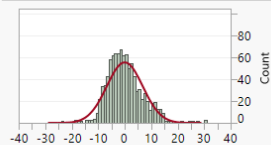 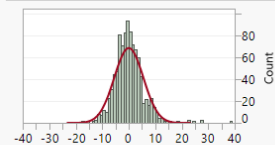

# Appendix E – Residual Distribution Analysis of Natural Gas Predictions

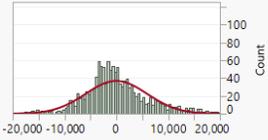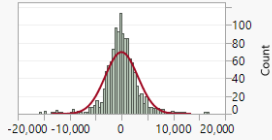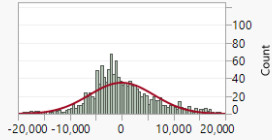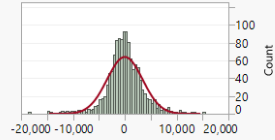
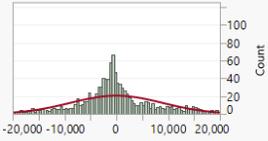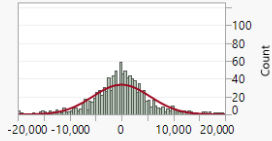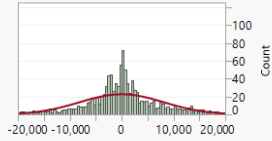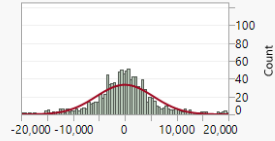
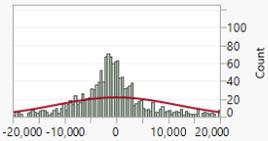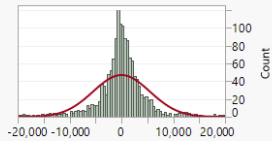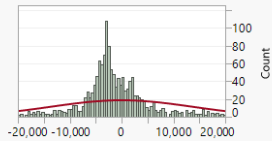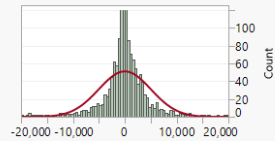
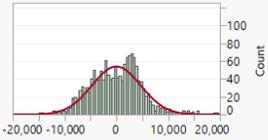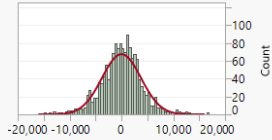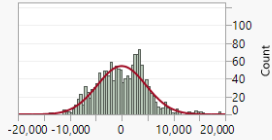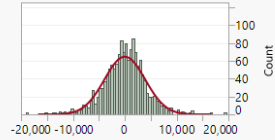
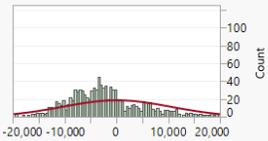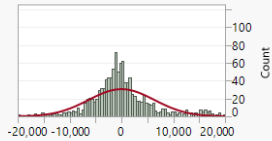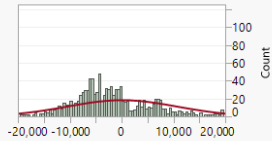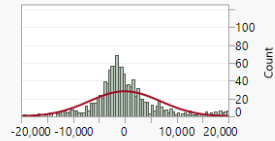

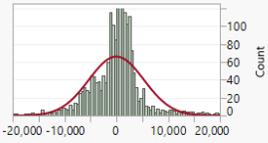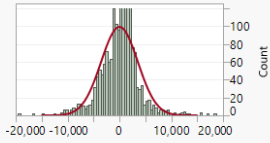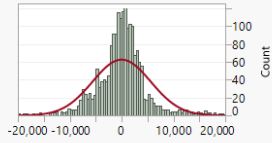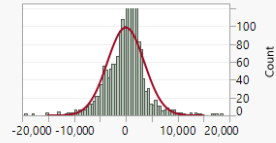
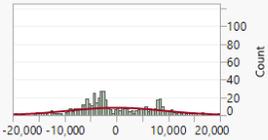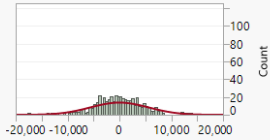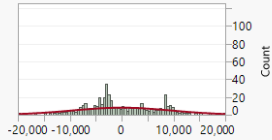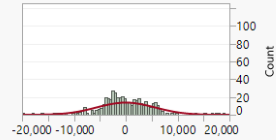
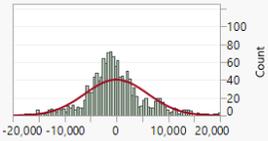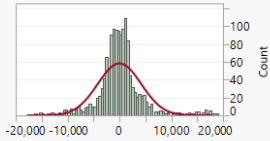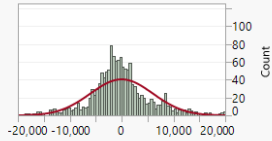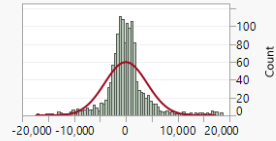
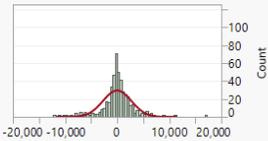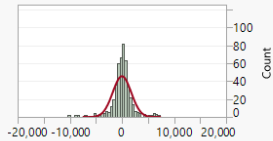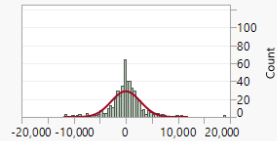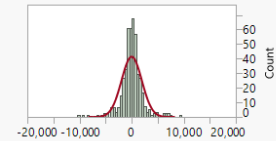
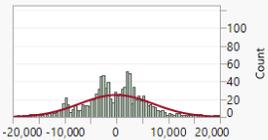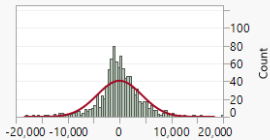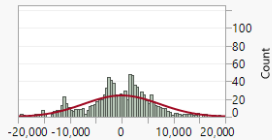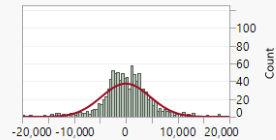

# Appendix F –Fluid Predictions and Actuals versus Days

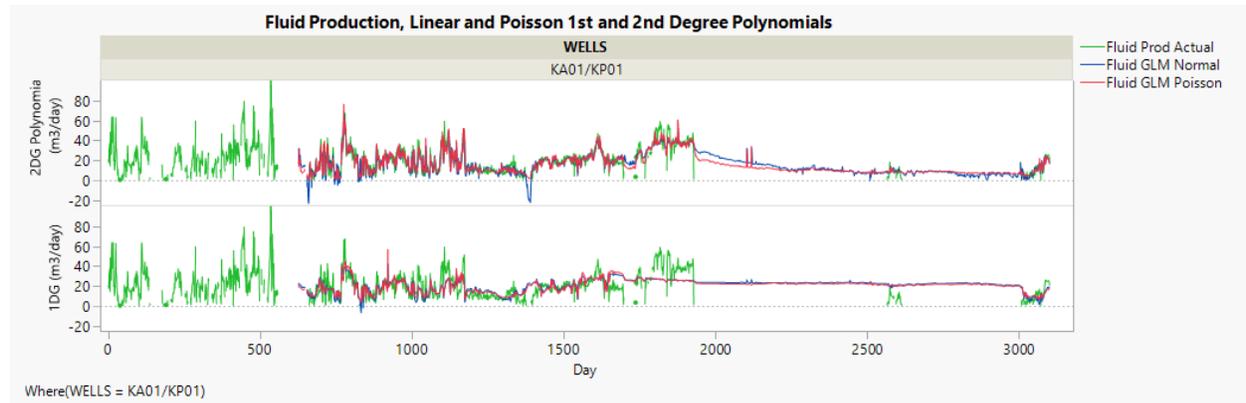

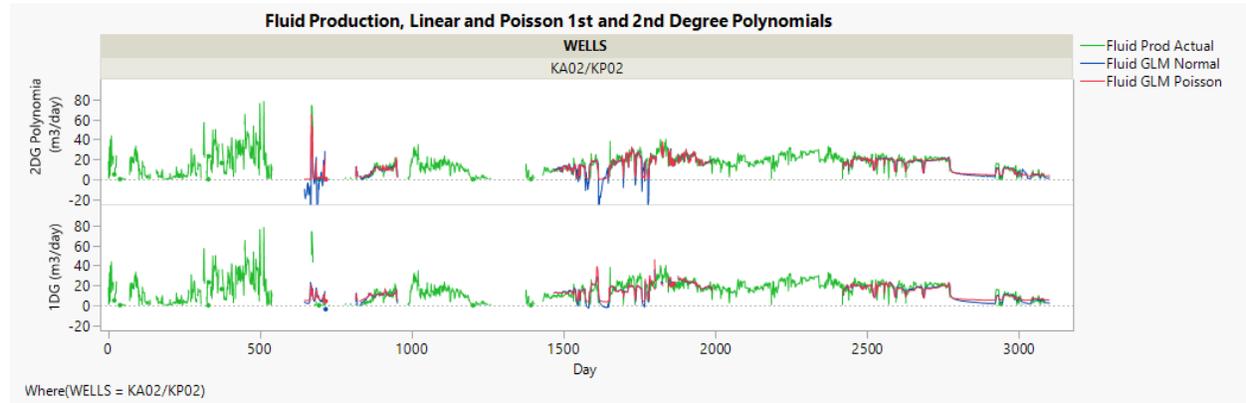

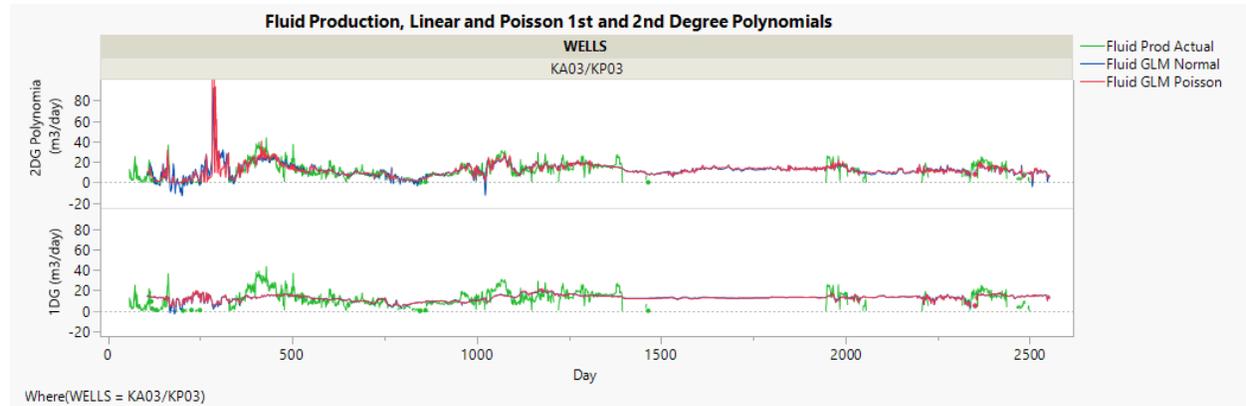

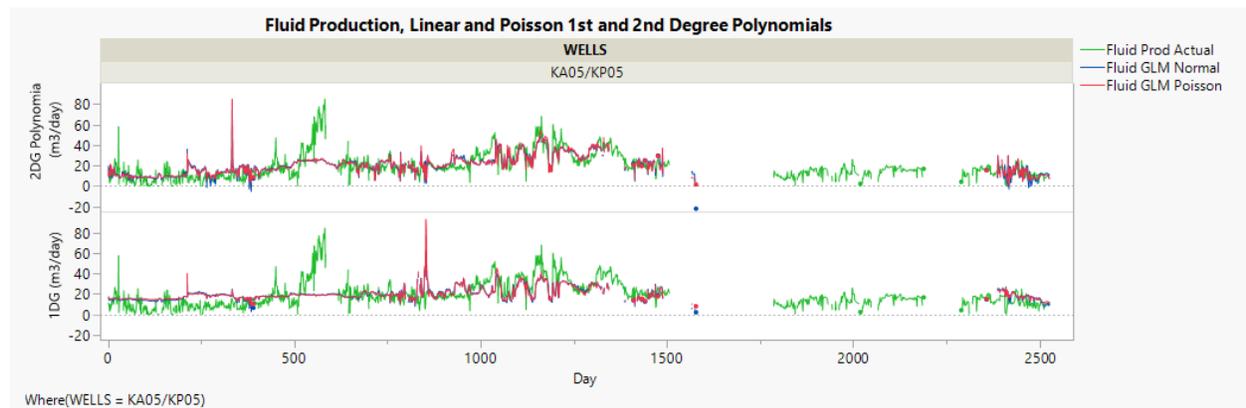

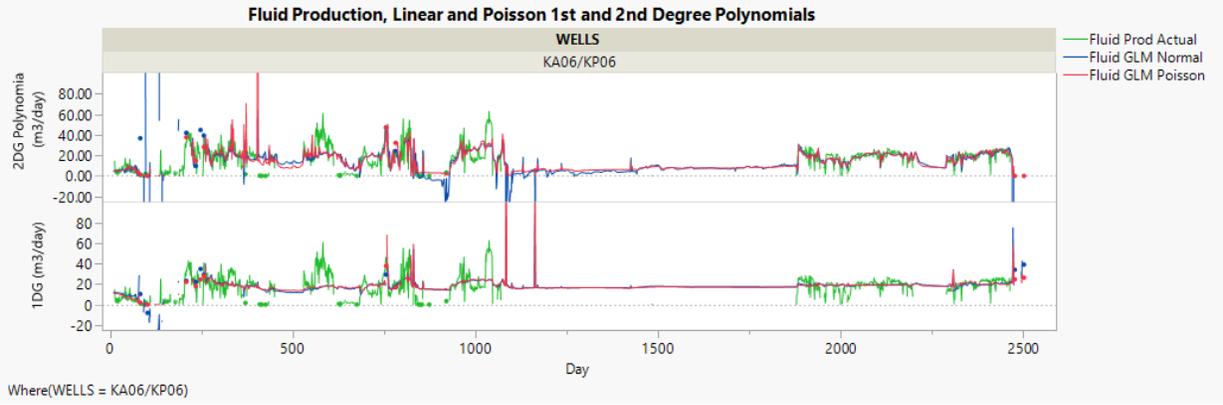
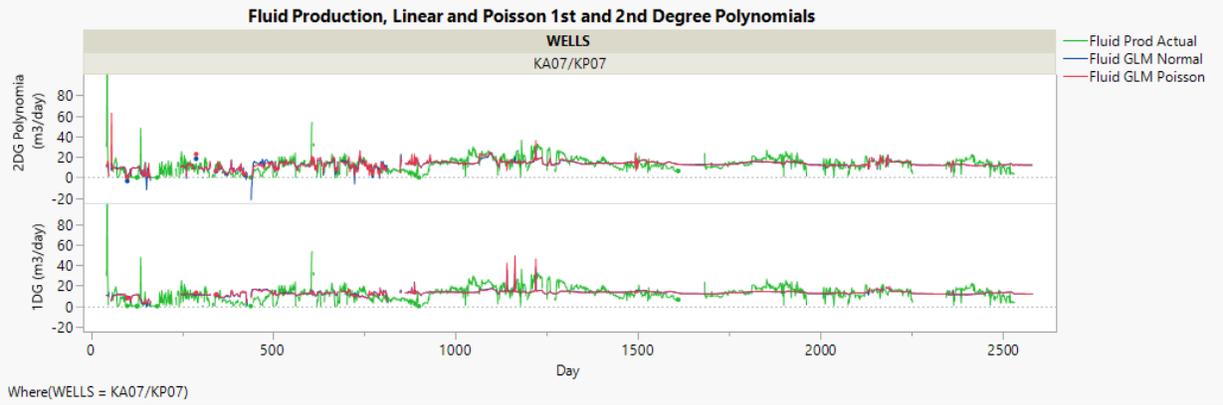
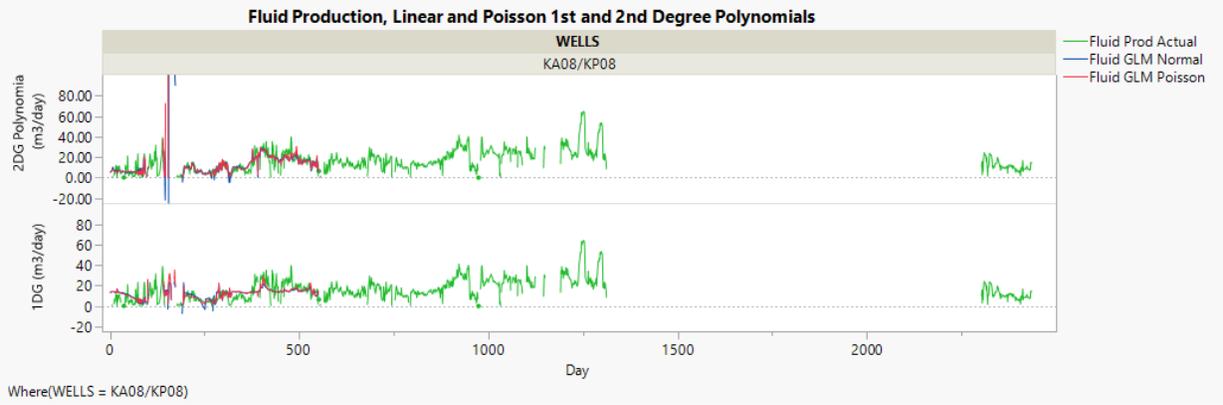
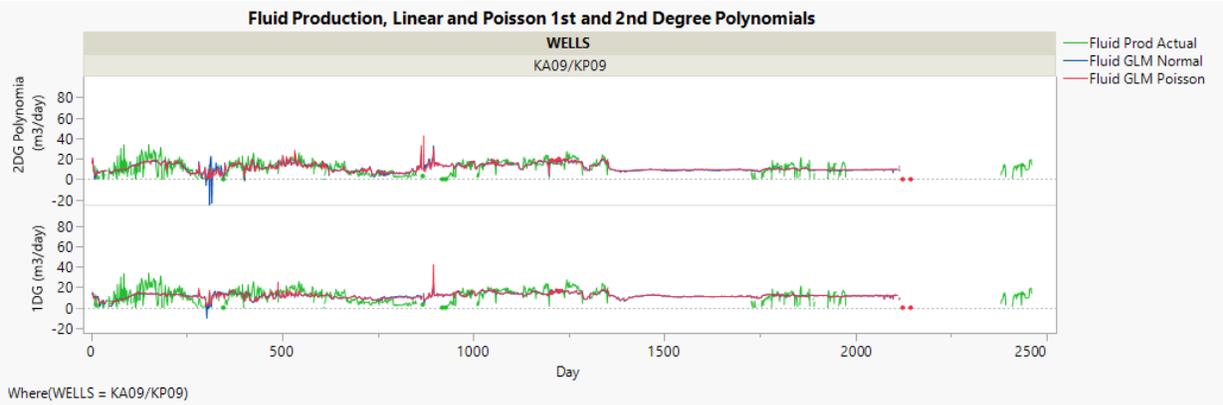

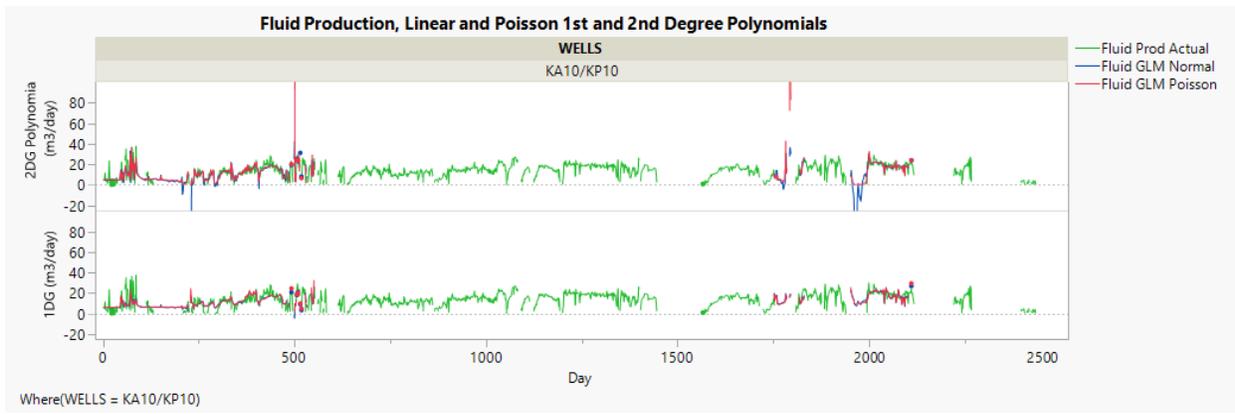

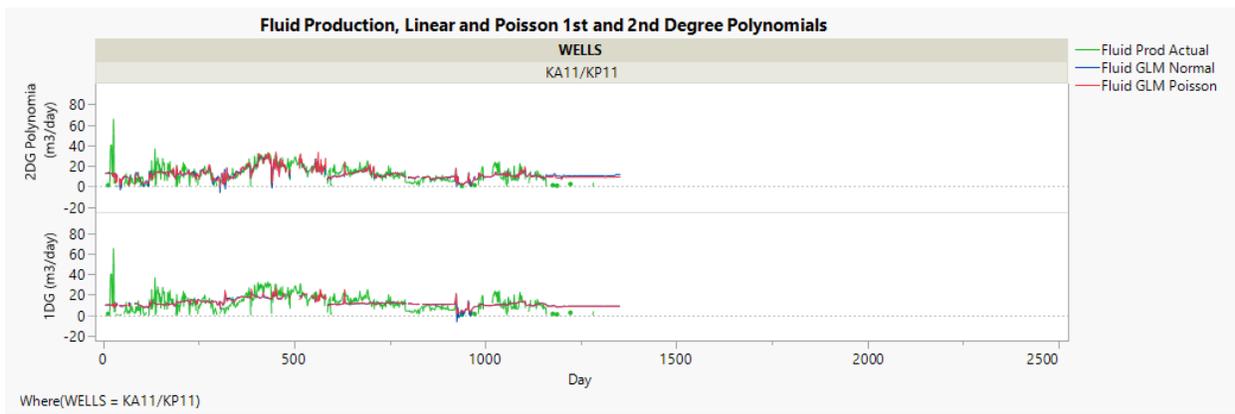

## Appendix G –Natural Gas Predictions and Actuals versus Days

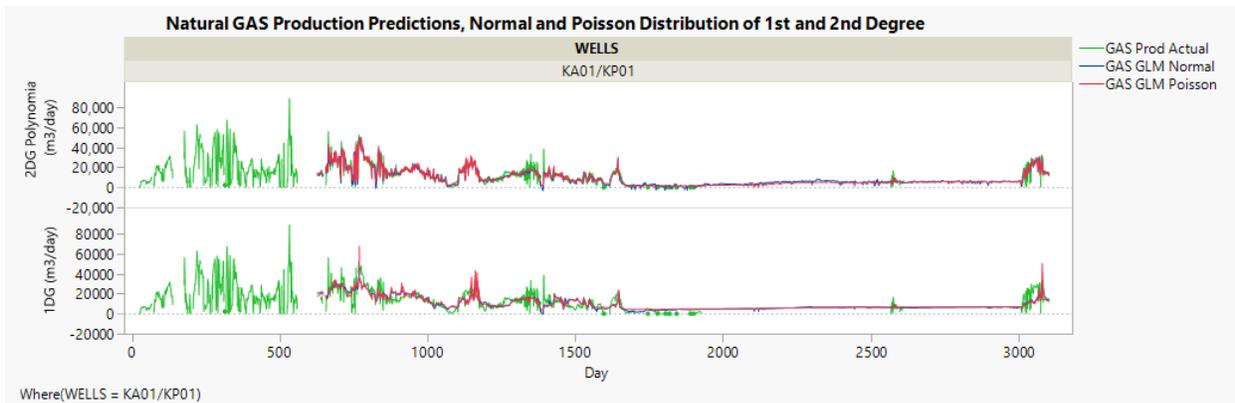

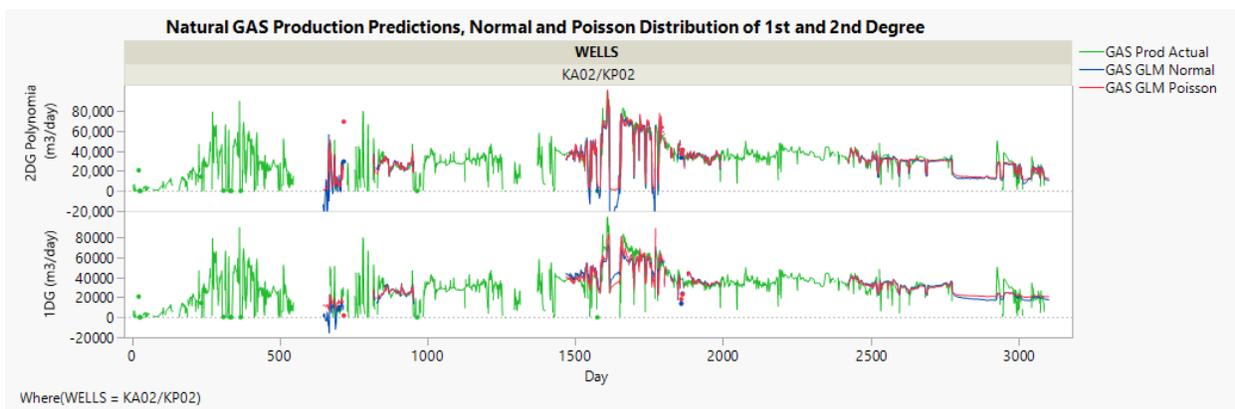

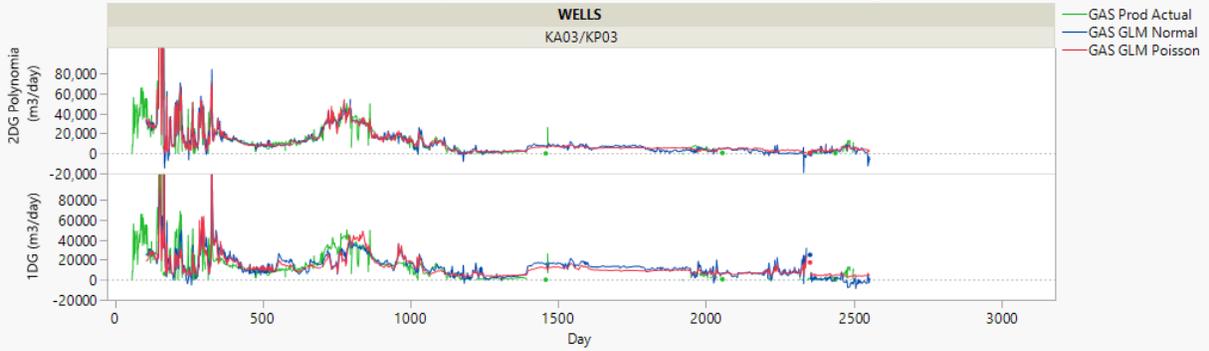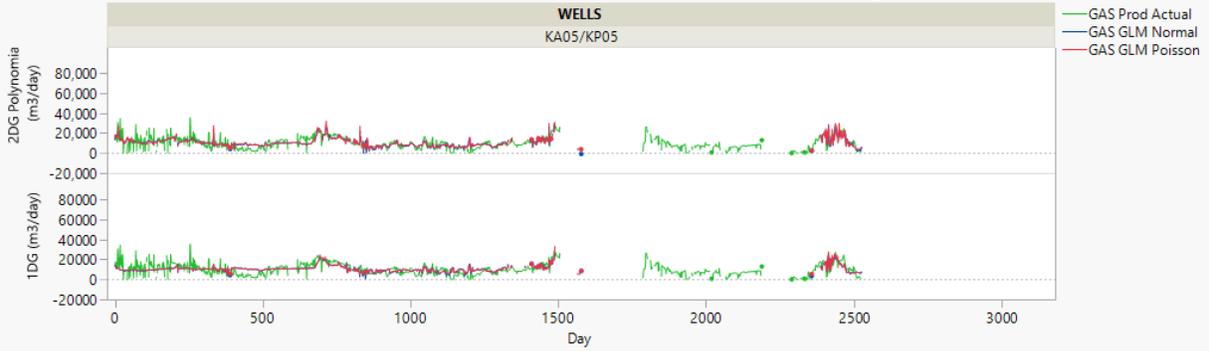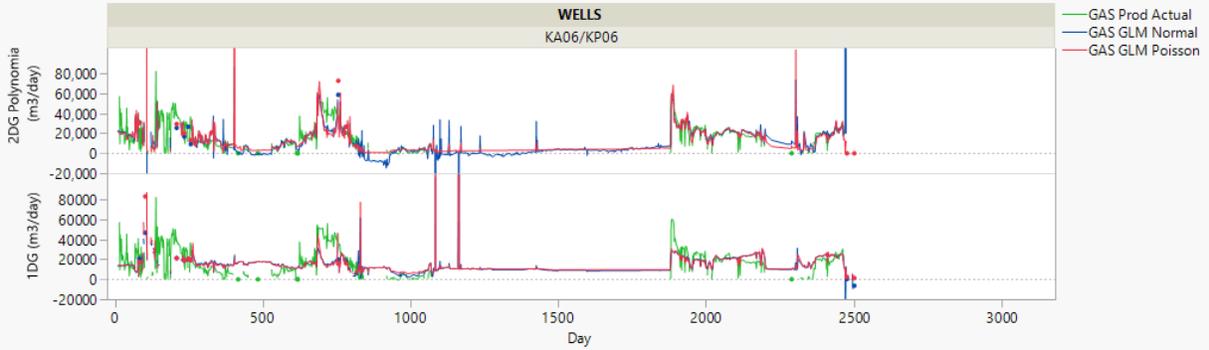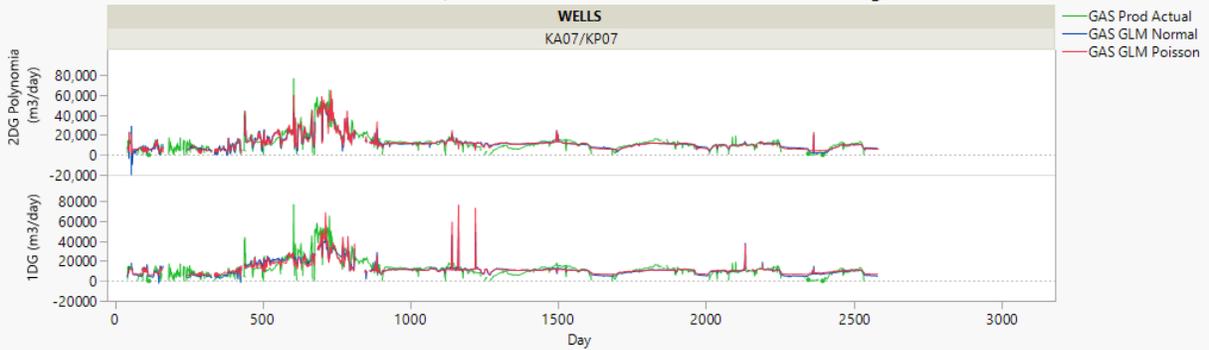

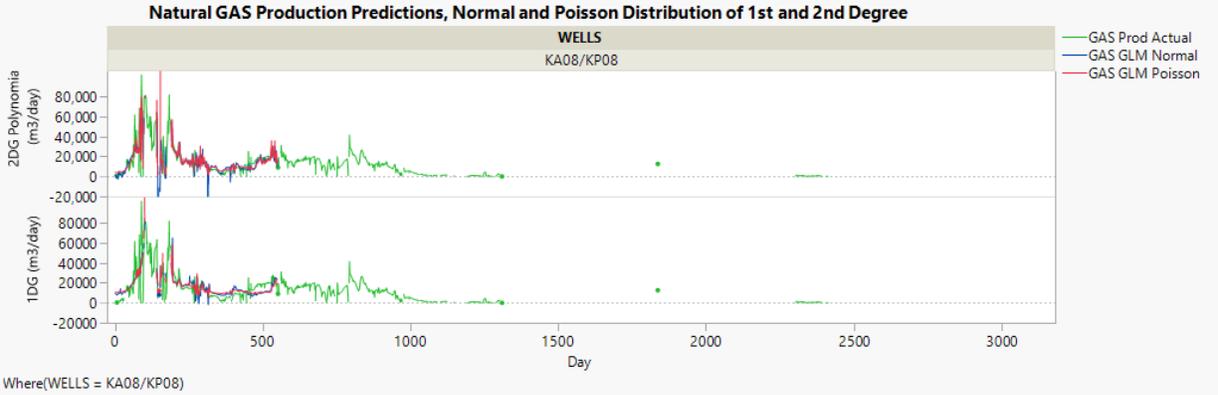

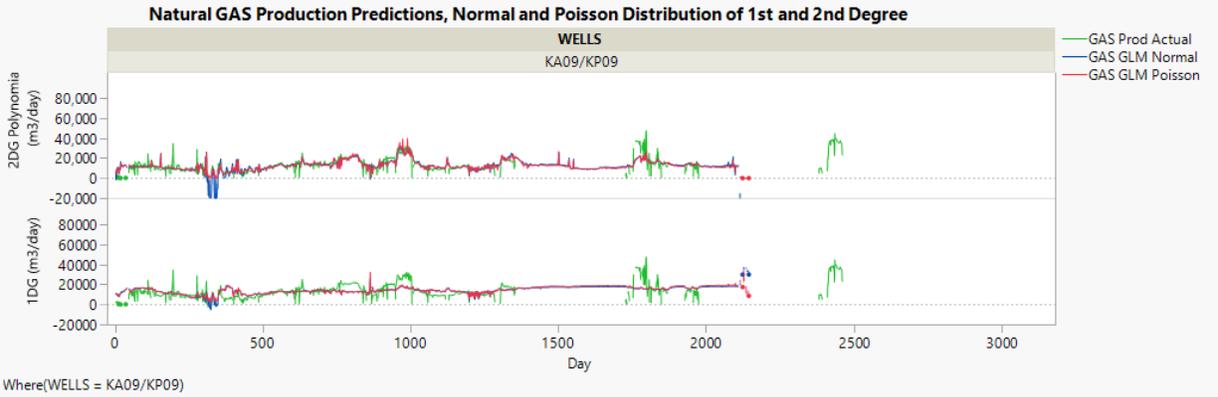

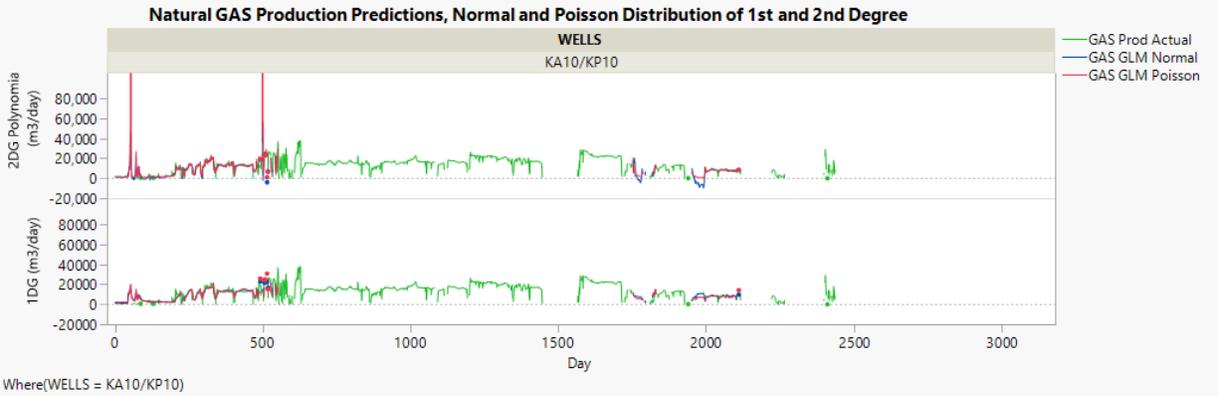

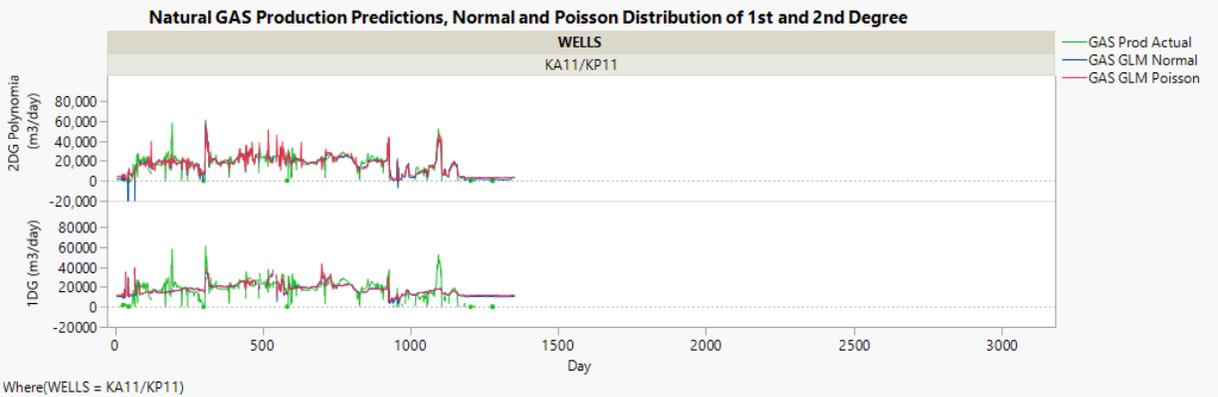

**Appendix H – Fluid Production Effect Summary for KP01, using Second Degree Quadratic Polynomials using Poisson Contribution**

| Source | LogWorth | PValue |
|---|---|---|
| THERMOCOUPLE 1 | 29.311 | 0.00000 |
| THERMOCOUPLE 11 | 18.124 | 0.00000 |
| THERMOCOUPLE 10 | 16.140 | 0.00000 |
| THERMOCOUPLE 8*THERMOCOUPLE 12 | 15.856 | 0.00000 |
| THERMOCOUPLE 1*THERMOCOUPLE 9 | 15.256 | 0.00000 |
| THERMOCOUPLE 1*THERMOCOUPLE 8 | 14.423 | 0.00000 |
| THERMOCOUPLE 6*THERMOCOUPLE 6 | 10.691 | 0.00000 |
| THERMOCOUPLE 7*THERMOCOUPLE 7 | 10.138 | 0.00000 |
| THERMOCOUPLE 12*THERMOCOUPLE 18 | 8.276 | 0.00000 |
| THERMOCOUPLE 13 | 7.543 | 0.00000 |
| THERMOCOUPLE 8*THERMOCOUPLE 9 | 6.984 | 0.00000 |
| THERMOCOUPLE 9*THERMOCOUPLE 11 | 6.906 | 0.00000 |
| THERMOCOUPLE 12*THERMOCOUPLE 17 | 6.798 | 0.00000 |
| THERMOCOUPLE 1*THERMOCOUPLE 16 | 6.701 | 0.00000 |
| THERMOCOUPLE 9*THERMOCOUPLE 9 | 6.593 | 0.00000 |
| THERMOCOUPLE 3*THERMOCOUPLE 12 | 6.573 | 0.00000 |
| THERMOCOUPLE 1*THERMOCOUPLE 20 | 6.431 | 0.00000 |
| THERMOCOUPLE 4*THERMOCOUPLE 18 | 6.376 | 0.00000 |
| THERMOCOUPLE 3*THERMOCOUPLE 11 | 6.150 | 0.00000 |
| THERMOCOUPLE 3*THERMOCOUPLE 17 | 5.812 | 0.00000 |
| THERMOCOUPLE 8*THERMOCOUPLE 14 | 5.565 | 0.00000 |
| THERMOCOUPLE 2 | 5.381 | 0.00000 |
| THERMOCOUPLE 12 | 5.334 | 0.00000 ^ |
| THERMOCOUPLE 3*THERMOCOUPLE 3 | 5.253 | 0.00001 |
| THERMOCOUPLE 10*THERMOCOUPLE 10 | 5.227 | 0.00001 |
| THERMOCOUPLE 16*THERMOCOUPLE 17 | 4.617 | 0.00002 |
| THERMOCOUPLE 2*THERMOCOUPLE 7 | 4.555 | 0.00003 |
| THERMOCOUPLE 4*THERMOCOUPLE 6 | 4.158 | 0.00007 |
| THERMOCOUPLE 2*THERMOCOUPLE 2 | 4.110 | 0.00008 |
| THERMOCOUPLE 1*THERMOCOUPLE 10 | 4.031 | 0.00009 |
| THERMOCOUPLE 4*THERMOCOUPLE 8 | 3.782 | 0.00017 |
| THERMOCOUPLE 3*THERMOCOUPLE 6 | 3.749 | 0.00018 |
| THERMOCOUPLE 2*THERMOCOUPLE 9 | 3.640 | 0.00023 |
| THERMOCOUPLE 2*THERMOCOUPLE 12 | 3.513 | 0.00031 |
| THERMOCOUPLE 10*THERMOCOUPLE 14 | 3.493 | 0.00032 |
| THERMOCOUPLE 1*THERMOCOUPLE 2 | 3.418 | 0.00038 |
| THERMOCOUPLE 4*THERMOCOUPLE 20 | 3.350 | 0.00045 |
| THERMOCOUPLE 9*THERMOCOUPLE 12 | 3.349 | 0.00045 |
| THERMOCOUPLE 2*THERMOCOUPLE 18 | 3.286 | 0.00052 |
| THERMOCOUPLE 3*THERMOCOUPLE 10 | 3.227 | 0.00059 |
| THERMOCOUPLE 12*THERMOCOUPLE 20 | 3.168 | 0.00068 |
| THERMOCOUPLE 16*THERMOCOUPLE 18 | 3.088 | 0.00082 |
| THERMOCOUPLE 2*THERMOCOUPLE 6 | 3.072 | 0.00085 |

| Source | LogWorth | PValue |
|---|---|---|
| THERMOCOUPLE 13*THERMOCOUPLE 13 | 2.986 | 0.00103 |
| THERMOCOUPLE 4*THERMOCOUPLE 13 | 2.877 | 0.00133 |
| THERMOCOUPLE 2*THERMOCOUPLE 11 | 2.855 | 0.00140 |
| THERMOCOUPLE 11*THERMOCOUPLE 12 | 2.805 | 0.00157 |
| THERMOCOUPLE 1*THERMOCOUPLE 17 | 2.796 | 0.00160 |
| THERMOCOUPLE 2*THERMOCOUPLE 10 | 2.743 | 0.00181 |
| THERMOCOUPLE 9*THERMOCOUPLE 17 | 2.662 | 0.00218 |
| THERMOCOUPLE 9*THERMOCOUPLE 13 | 2.640 | 0.00229 |
| THERMOCOUPLE 1*THERMOCOUPLE 3 | 2.616 | 0.00242 |
| THERMOCOUPLE 6*THERMOCOUPLE 18 | 2.606 | 0.00248 |
| THERMOCOUPLE 7*THERMOCOUPLE 17 | 2.558 | 0.00277 |
| THERMOCOUPLE 17*THERMOCOUPLE 17 | 2.482 | 0.00330 |
| THERMOCOUPLE 2*THERMOCOUPLE 14 | 2.459 | 0.00348 |
| THERMOCOUPLE 1*THERMOCOUPLE 4 | 2.445 | 0.00359 |
| THERMOCOUPLE 8*THERMOCOUPLE 10 | 2.357 | 0.00439 |
| THERMOCOUPLE 1*THERMOCOUPLE 1 | 2.349 | 0.00448 |
| THERMOCOUPLE 13*THERMOCOUPLE 17 | 2.318 | 0.00480 |
| THERMOCOUPLE 6*THERMOCOUPLE 17 | 2.273 | 0.00533 |
| THERMOCOUPLE 11*THERMOCOUPLE 13 | 2.247 | 0.00566 |
| THERMOCOUPLE 3*THERMOCOUPLE 16 | 2.240 | 0.00576 |
| THERMOCOUPLE 13*THERMOCOUPLE 14 | 2.216 | 0.00609 |
| THERMOCOUPLE 10*THERMOCOUPLE 12 | 2.197 | 0.00636 |
| THERMOCOUPLE 6*THERMOCOUPLE 8 | 2.192 | 0.00643 |
| THERMOCOUPLE 8*THERMOCOUPLE 11 | 2.176 | 0.00667 |
| THERMOCOUPLE 11*THERMOCOUPLE 20 | 2.171 | 0.00674 |
| THERMOCOUPLE 6*THERMOCOUPLE 12 | 2.164 | 0.00686 |
| THERMOCOUPLE 2*THERMOCOUPLE 17 | 2.149 | 0.00710 |
| THERMOCOUPLE 3*THERMOCOUPLE 20 | 2.131 | 0.00739 |
| THERMOCOUPLE 7*THERMOCOUPLE 16 | 2.123 | 0.00754 |
| THERMOCOUPLE 7*THERMOCOUPLE 13 | 2.083 | 0.00826 |
| THERMOCOUPLE 4*THERMOCOUPLE 14 | 2.074 | 0.00843 |
| THERMOCOUPLE 8*THERMOCOUPLE 8 | 2.028 | 0.00938 |
| THERMOCOUPLE 1*THERMOCOUPLE 11 | 2.002 | 0.00996 |
| THERMOCOUPLE 4*THERMOCOUPLE 4 | 1.990 | 0.01024 |
| THERMOCOUPLE 4*THERMOCOUPLE 16 | 1.936 | 0.01160 |
| THERMOCOUPLE 8*THERMOCOUPLE 13 | 1.933 | 0.01168 |
| THERMOCOUPLE 4*THERMOCOUPLE 9 | 1.897 | 0.01267 |
| THERMOCOUPLE 7*THERMOCOUPLE 8 | 1.854 | 0.01401 |
| THERMOCOUPLE 9*THERMOCOUPLE 16 | 1.823 | 0.01502 |
| THERMOCOUPLE 6*THERMOCOUPLE 7 | 1.789 | 0.01625 |
| THERMOCOUPLE 1*THERMOCOUPLE 7 | 1.788 | 0.01631 |
| THERMOCOUPLE 6*THERMOCOUPLE 14 | 1.778 | 0.01666 |
| THERMOCOUPLE 6*THERMOCOUPLE 11 | 1.751 | 0.01772 |
| THERMOCOUPLE 4*THERMOCOUPLE 10 | 1.721 | 0.01903 |
| THERMOCOUPLE 7 | 1.697 | 0.02011 ^ |

| Source | LogWorth | PValue |
|---|---|---|
| THERMOCOUPLE 2*THERMOCOUPLE 20 | 1.599 | 0.02516 |
| THERMOCOUPLE 8 | 1.491 | 0.03229 ^ |
| THERMOCOUPLE 1*THERMOCOUPLE 18 | 1.482 | 0.03298 |
| THERMOCOUPLE 7*THERMOCOUPLE 9 | 1.415 | 0.03843 |
| THERMOCOUPLE 14*THERMOCOUPLE 18 | 1.410 | 0.03893 |
| THERMOCOUPLE 4 | 1.389 | 0.04081 ^ |
| THERMOCOUPLE 10*THERMOCOUPLE 16 | 1.363 | 0.04340 |
| THERMOCOUPLE 13*THERMOCOUPLE 16 | 1.341 | 0.04558 |
| THERMOCOUPLE 2*THERMOCOUPLE 4 | 1.324 | 0.04743 |
| THERMOCOUPLE 10*THERMOCOUPLE 11 | 1.321 | 0.04770 |
| THERMOCOUPLE 9*THERMOCOUPLE 18 | 1.303 | 0.04974 |
| THERMOCOUPLE 6*THERMOCOUPLE 13 | 1.201 | 0.06291 |
| THERMOCOUPLE 9*THERMOCOUPLE 14 | 1.197 | 0.06350 |
| THERMOCOUPLE 12*THERMOCOUPLE 14 | 1.150 | 0.07085 |
| THERMOCOUPLE 4*THERMOCOUPLE 12 | 1.113 | 0.07716 |
| THERMOCOUPLE 7*THERMOCOUPLE 14 | 1.106 | 0.07827 |
| THERMOCOUPLE 7*THERMOCOUPLE 20 | 1.078 | 0.08352 |
| THERMOCOUPLE 13*THERMOCOUPLE 18 | 1.052 | 0.08875 |
| THERMOCOUPLE 16*THERMOCOUPLE 16 | 1.038 | 0.09154 |
| THERMOCOUPLE 3*THERMOCOUPLE 4 | 1.022 | 0.09505 |
| THERMOCOUPLE 14*THERMOCOUPLE 16 | 1.013 | 0.09708 |
| THERMOCOUPLE 14*THERMOCOUPLE 14 | 1.010 | 0.09772 |
| THERMOCOUPLE 6*THERMOCOUPLE 20 | 0.981 | 0.10438 |
| THERMOCOUPLE 8*THERMOCOUPLE 17 | 0.978 | 0.10510 |
| THERMOCOUPLE 17 | 0.961 | 0.10934 ^ |
| THERMOCOUPLE 4*THERMOCOUPLE 17 | 0.946 | 0.11324 |
| THERMOCOUPLE 10*THERMOCOUPLE 13 | 0.943 | 0.11396 |
| THERMOCOUPLE 11*THERMOCOUPLE 18 | 0.931 | 0.11728 |
| THERMOCOUPLE 2*THERMOCOUPLE 16 | 0.905 | 0.12452 |
| THERMOCOUPLE 7*THERMOCOUPLE 10 | 0.893 | 0.12787 |
| THERMOCOUPLE 17*THERMOCOUPLE 18 | 0.865 | 0.13657 |
| THERMOCOUPLE 18 | 0.850 | 0.14111 ^ |
| THERMOCOUPLE 12*THERMOCOUPLE 13 | 0.846 | 0.14264 |
| THERMOCOUPLE 9*THERMOCOUPLE 10 | 0.835 | 0.14623 |
| THERMOCOUPLE 8*THERMOCOUPLE 18 | 0.761 | 0.17345 |
| THERMOCOUPLE 3*THERMOCOUPLE 8 | 0.741 | 0.18141 |
| THERMOCOUPLE 3*THERMOCOUPLE 9 | 0.704 | 0.19792 |
| THERMOCOUPLE 4*THERMOCOUPLE 11 | 0.701 | 0.19913 |
| THERMOCOUPLE 6 | 0.688 | 0.20491 ^ |
| THERMOCOUPLE 12*THERMOCOUPLE 16 | 0.673 | 0.21226 |
| THERMOCOUPLE 3*THERMOCOUPLE 18 | 0.655 | 0.22123 |
| THERMOCOUPLE 7*THERMOCOUPLE 11 | 0.650 | 0.22368 |
| THERMOCOUPLE 2*THERMOCOUPLE 13 | 0.645 | 0.22643 |
| THERMOCOUPLE 7*THERMOCOUPLE 12 | 0.622 | 0.23876 |
| THERMOCOUPLE 4*THERMOCOUPLE 7 | 0.608 | 0.24651 |

| Source | LogWorth | PValue |
|---|---|---|
| THERMOCOUPLE 16 | 0.557 | 0.27709 ^ |
| THERMOCOUPLE 10*THERMOCOUPLE 17 | 0.533 | 0.29318 |
| THERMOCOUPLE 3 | 0.486 | 0.32629 ^ |
| THERMOCOUPLE 14 | 0.483 | 0.32879 ^ |
| THERMOCOUPLE 8*THERMOCOUPLE 20 | 0.477 | 0.33332 |
| THERMOCOUPLE 8*THERMOCOUPLE 16 | 0.459 | 0.34780 |
| THERMOCOUPLE 6*THERMOCOUPLE 9 | 0.386 | 0.41136 |
| THERMOCOUPLE 10*THERMOCOUPLE 18 | 0.384 | 0.41296 |
| THERMOCOUPLE 2*THERMOCOUPLE 8 | 0.377 | 0.42006 |
| THERMOCOUPLE 17*THERMOCOUPLE 20 | 0.370 | 0.42644 |
| THERMOCOUPLE 20 | 0.370 | 0.42678 ^ |
| THERMOCOUPLE 1*THERMOCOUPLE 12 | 0.351 | 0.44536 |
| THERMOCOUPLE 9 | 0.350 | 0.44719 ^ |
| THERMOCOUPLE 14*THERMOCOUPLE 20 | 0.324 | 0.47384 |
| THERMOCOUPLE 13*THERMOCOUPLE 20 | 0.290 | 0.51249 |
| THERMOCOUPLE 11*THERMOCOUPLE 16 | 0.289 | 0.51422 |
| THERMOCOUPLE 3*THERMOCOUPLE 13 | 0.283 | 0.52093 |
| THERMOCOUPLE 9*THERMOCOUPLE 20 | 0.282 | 0.52204 |
| THERMOCOUPLE 3*THERMOCOUPLE 14 | 0.277 | 0.52801 |
| THERMOCOUPLE 7*THERMOCOUPLE 18 | 0.273 | 0.53348 |
| THERMOCOUPLE 1*THERMOCOUPLE 6 | 0.251 | 0.56121 |
| THERMOCOUPLE 6*THERMOCOUPLE 10 | 0.234 | 0.58373 |
| THERMOCOUPLE 1*THERMOCOUPLE 13 | 0.227 | 0.59335 |
| THERMOCOUPLE 11*THERMOCOUPLE 17 | 0.194 | 0.63957 |
| THERMOCOUPLE 11*THERMOCOUPLE 14 | 0.186 | 0.65203 |
| THERMOCOUPLE 14*THERMOCOUPLE 17 | 0.177 | 0.66528 |
| THERMOCOUPLE 6*THERMOCOUPLE 16 | 0.173 | 0.67071 |
| THERMOCOUPLE 3*THERMOCOUPLE 7 | 0.153 | 0.70268 |
| THERMOCOUPLE 18*THERMOCOUPLE 20 | 0.127 | 0.74603 |
| THERMOCOUPLE 2*THERMOCOUPLE 3 | 0.115 | 0.76684 |
| THERMOCOUPLE 18*THERMOCOUPLE 18 | 0.098 | 0.79712 |
| THERMOCOUPLE 16*THERMOCOUPLE 20 | 0.082 | 0.82774 |
| THERMOCOUPLE 12*THERMOCOUPLE 12 | 0.060 | 0.87142 |
| THERMOCOUPLE 1*THERMOCOUPLE 14 | 0.046 | 0.89915 |
| THERMOCOUPLE 10*THERMOCOUPLE 20 | 0.034 | 0.92496 |
| THERMOCOUPLE 11*THERMOCOUPLE 11 | 0.022 | 0.95064 |
| THERMOCOUPLE 20*THERMOCOUPLE 20 | 0.012 | 0.97227 |

**Appendix I – Natural Gas Production Effect Summary for KP01, using Second Degree Quadratic Polynomials using Poisson Contribution**

| Source | LogWorth | PValue |
|---|---|---|
| THERMOCOUPLE 10*THERMOCOUPLE 12 | 14192.84 | 0.00000 |
| THERMOCOUPLE 10*THERMOCOUPLE 13 | 12668.21 | 0.00000 |
| THERMOCOUPLE 11*THERMOCOUPLE 12 | 9948.348 | 0.00000 |

| Source | LogWorth | PValue |
|---|---|---|
| THERMOCOUPLE 14 | 9040.609 | 0.00000 |
| THERMOCOUPLE 10*THERMOCOUPLE 14 | 7235.084 | 0.00000 |
| THERMOCOUPLE 13 | 6822.531 | 0.00000 ^ |
| THERMOCOUPLE 12*THERMOCOUPLE 20 | 6646.944 | 0.00000 |
| THERMOCOUPLE 3 | 6041.172 | 0.00000 |
| THERMOCOUPLE 8*THERMOCOUPLE 13 | 5462.259 | 0.00000 |
| THERMOCOUPLE 11*THERMOCOUPLE 20 | 5151.365 | 0.00000 |
| THERMOCOUPLE 10*THERMOCOUPLE 16 | 5043.121 | 0.00000 |
| THERMOCOUPLE 8*THERMOCOUPLE 17 | 4964.610 | 0.00000 |
| THERMOCOUPLE 3*THERMOCOUPLE 11 | 4876.550 | 0.00000 |
| THERMOCOUPLE 11*THERMOCOUPLE 16 | 4775.787 | 0.00000 |
| THERMOCOUPLE 11*THERMOCOUPLE 14 | 4678.654 | 0.00000 |
| THERMOCOUPLE 11*THERMOCOUPLE 13 | 4090.118 | 0.00000 |
| THERMOCOUPLE 12*THERMOCOUPLE 18 | 3954.697 | 0.00000 |
| THERMOCOUPLE 17 | 3549.411 | 0.00000 ^ |
| THERMOCOUPLE 8*THERMOCOUPLE 18 | 3400.160 | 0.00000 |
| THERMOCOUPLE 12 | 3383.417 | 0.00000 ^ |
| THERMOCOUPLE 9*THERMOCOUPLE 16 | 3333.171 | 0.00000 |
| THERMOCOUPLE 9*THERMOCOUPLE 20 | 3181.944 | 0.00000 |
| THERMOCOUPLE 16*THERMOCOUPLE 18 | 3143.348 | 0.00000 |
| THERMOCOUPLE 17*THERMOCOUPLE 18 | 3075.458 | 0.00000 |
| THERMOCOUPLE 10*THERMOCOUPLE 20 | 3061.580 | 0.00000 |
| THERMOCOUPLE 2*THERMOCOUPLE 18 | 2933.734 | 0.00000 |
| THERMOCOUPLE 14*THERMOCOUPLE 20 | 2792.498 | 0.00000 |
| THERMOCOUPLE 9*THERMOCOUPLE 18 | 2694.560 | 0.00000 |
| THERMOCOUPLE 16*THERMOCOUPLE 20 | 2667.985 | 0.00000 |
| THERMOCOUPLE 14*THERMOCOUPLE 18 | 2531.732 | 0.00000 |
| THERMOCOUPLE 9*THERMOCOUPLE 10 | 2499.602 | 0.00000 |
| THERMOCOUPLE 1*THERMOCOUPLE 8 | 2335.276 | 0.00000 |
| THERMOCOUPLE 17*THERMOCOUPLE 20 | 2287.560 | 0.00000 |
| THERMOCOUPLE 2*THERMOCOUPLE 16 | 2219.030 | 0.00000 |
| THERMOCOUPLE 13*THERMOCOUPLE 20 | 2194.403 | 0.00000 |
| THERMOCOUPLE 9*THERMOCOUPLE 11 | 2181.372 | 0.00000 |
| THERMOCOUPLE 11*THERMOCOUPLE 18 | 1870.418 | 0.00000 |
| THERMOCOUPLE 7*THERMOCOUPLE 9 | 1840.774 | 0.00000 |
| THERMOCOUPLE 10*THERMOCOUPLE 17 | 1818.360 | 0.00000 |
| THERMOCOUPLE 3*THERMOCOUPLE 17 | 1791.216 | 0.00000 |
| THERMOCOUPLE 8*THERMOCOUPLE 9 | 1769.834 | 0.00000 |
| THERMOCOUPLE 2*THERMOCOUPLE 20 | 1733.780 | 0.00000 |
| THERMOCOUPLE 3*THERMOCOUPLE 12 | 1724.504 | 0.00000 |
| THERMOCOUPLE 8 | 1723.621 | 0.00000 ^ |
| THERMOCOUPLE 8*THERMOCOUPLE 14 | 1720.777 | 0.00000 |
| THERMOCOUPLE 4*THERMOCOUPLE 11 | 1691.791 | 0.00000 |
| THERMOCOUPLE 3*THERMOCOUPLE 10 | 1681.947 | 0.00000 |
| THERMOCOUPLE 3*THERMOCOUPLE 18 | 1519.351 | 0.00000 |

| Source | LogWorth | PValue |
|---|---|---|
| THERMOCOUPLE 9*THERMOCOUPLE 13 | 1480.616 | 0.00000 |
| THERMOCOUPLE 13*THERMOCOUPLE 18 | 1436.908 | 0.00000 |
| THERMOCOUPLE 4*THERMOCOUPLE 16 | 1352.723 | 0.00000 |
| THERMOCOUPLE 16 | 1291.750 | 0.00000 ^ |
| THERMOCOUPLE 7*THERMOCOUPLE 13 | 1287.174 | 0.00000 |
| THERMOCOUPLE 8*THERMOCOUPLE 10 | 1250.464 | 0.00000 |
| THERMOCOUPLE 12*THERMOCOUPLE 17 | 1224.530 | 0.00000 |
| THERMOCOUPLE 4*THERMOCOUPLE 10 | 1191.125 | 0.00000 |
| THERMOCOUPLE 6*THERMOCOUPLE 8 | 1108.723 | 0.00000 |
| THERMOCOUPLE 2*THERMOCOUPLE 10 | 1094.295 | 0.00000 |
| THERMOCOUPLE 14*THERMOCOUPLE 17 | 1092.778 | 0.00000 |
| THERMOCOUPLE 10*THERMOCOUPLE 11 | 1088.394 | 0.00000 |
| THERMOCOUPLE 7*THERMOCOUPLE 18 | 987.248 | 0.00000 |
| THERMOCOUPLE 16*THERMOCOUPLE 17 | 949.156 | 0.00000 |
| THERMOCOUPLE 4*THERMOCOUPLE 9 | 901.515 | 0.00000 |
| THERMOCOUPLE 1*THERMOCOUPLE 9 | 886.478 | 0.00000 |
| THERMOCOUPLE 12*THERMOCOUPLE 16 | 881.735 | 0.00000 |
| THERMOCOUPLE 1*THERMOCOUPLE 10 | 864.574 | 0.00000 |
| THERMOCOUPLE 8*THERMOCOUPLE 12 | 856.904 | 0.00000 |
| THERMOCOUPLE 4*THERMOCOUPLE 8 | 855.778 | 0.00000 |
| THERMOCOUPLE 18*THERMOCOUPLE 20 | 854.751 | 0.00000 |
| THERMOCOUPLE 11 | 806.005 | 0.00000 ^ |
| THERMOCOUPLE 7*THERMOCOUPLE 10 | 697.744 | 0.00000 |
| THERMOCOUPLE 6*THERMOCOUPLE 7 | 677.165 | 0.00000 |
| THERMOCOUPLE 12*THERMOCOUPLE 14 | 672.055 | 0.00000 |
| THERMOCOUPLE 2*THERMOCOUPLE 13 | 654.229 | 0.00000 |
| THERMOCOUPLE 7 | 621.852 | 0.00000 ^ |
| THERMOCOUPLE 4*THERMOCOUPLE 13 | 586.256 | 0.00000 |
| THERMOCOUPLE 2 | 563.999 | 0.00000 ^ |
| THERMOCOUPLE 7*THERMOCOUPLE 20 | 547.532 | 0.00000 |
| THERMOCOUPLE 4 | 539.084 | 0.00000 ^ |
| THERMOCOUPLE 20 | 505.626 | 0.00000 ^ |
| THERMOCOUPLE 6 | 494.842 | 0.00000 ^ |
| THERMOCOUPLE 8*THERMOCOUPLE 16 | 494.738 | 0.00000 |
| THERMOCOUPLE 9*THERMOCOUPLE 14 | 489.749 | 0.00000 |
| THERMOCOUPLE 6*THERMOCOUPLE 12 | 466.785 | 0.00000 |
| THERMOCOUPLE 7*THERMOCOUPLE 16 | 463.291 | 0.00000 |
| THERMOCOUPLE 13*THERMOCOUPLE 17 | 457.124 | 0.00000 |
| THERMOCOUPLE 4*THERMOCOUPLE 18 | 454.675 | 0.00000 |
| THERMOCOUPLE 9 | 406.152 | 0.00000 ^ |
| THERMOCOUPLE 4*THERMOCOUPLE 14 | 397.978 | 0.00000 |
| THERMOCOUPLE 7*THERMOCOUPLE 12 | 396.837 | 0.00000 |
| THERMOCOUPLE 6*THERMOCOUPLE 20 | 370.049 | 0.00000 |
| THERMOCOUPLE 7*THERMOCOUPLE 14 | 345.195 | 0.00000 |
| THERMOCOUPLE 2*THERMOCOUPLE 11 | 318.887 | 0.00000 |

| Source | LogWorth | PValue |
|---|---|---|
| THERMOCOUPLE 2*THERMOCOUPLE 12 | 302.629 | 0.00000 |
| THERMOCOUPLE 2*THERMOCOUPLE 14 | 301.424 | 0.00000 |
| THERMOCOUPLE 1*THERMOCOUPLE 6 | 291.338 | 0.00000 |
| THERMOCOUPLE 6*THERMOCOUPLE 9 | 264.373 | 0.00000 |
| THERMOCOUPLE 13*THERMOCOUPLE 14 | 252.986 | 0.00000 |
| THERMOCOUPLE 3*THERMOCOUPLE 6 | 226.682 | 0.00000 |
| THERMOCOUPLE 6*THERMOCOUPLE 14 | 206.929 | 0.00000 |
| THERMOCOUPLE 10*THERMOCOUPLE 18 | 192.160 | 0.00000 |
| THERMOCOUPLE 18 | 188.635 | 0.00000 ^ |
| THERMOCOUPLE 2*THERMOCOUPLE 8 | 182.524 | 0.00000 |
| THERMOCOUPLE 14*THERMOCOUPLE 16 | 164.603 | 0.00000 |
| THERMOCOUPLE 3*THERMOCOUPLE 20 | 157.679 | 0.00000 |
| THERMOCOUPLE 4*THERMOCOUPLE 7 | 153.898 | 0.00000 |
| THERMOCOUPLE 6*THERMOCOUPLE 16 | 150.971 | 0.00000 |
| THERMOCOUPLE 4*THERMOCOUPLE 20 | 149.998 | 0.00000 |
| THERMOCOUPLE 1*THERMOCOUPLE 17 | 144.575 | 0.00000 |
| THERMOCOUPLE 1*THERMOCOUPLE 12 | 129.048 | 0.00000 |
| THERMOCOUPLE 2*THERMOCOUPLE 17 | 125.942 | 0.00000 |
| THERMOCOUPLE 4*THERMOCOUPLE 12 | 118.599 | 0.00000 |
| THERMOCOUPLE 8*THERMOCOUPLE 20 | 114.104 | 0.00000 |
| THERMOCOUPLE 1*THERMOCOUPLE 4 | 111.391 | 0.00000 |
| THERMOCOUPLE 3*THERMOCOUPLE 14 | 108.140 | 0.00000 |
| THERMOCOUPLE 1*THERMOCOUPLE 20 | 106.870 | 0.00000 |
| THERMOCOUPLE 8*THERMOCOUPLE 11 | 105.829 | 0.00000 |
| THERMOCOUPLE 12*THERMOCOUPLE 13 | 105.768 | 0.00000 |
| THERMOCOUPLE 10 | 78.680 | 0.00000 ^ |
| THERMOCOUPLE 3*THERMOCOUPLE 9 | 76.370 | 0.00000 |
| THERMOCOUPLE 13*THERMOCOUPLE 16 | 75.562 | 0.00000 |
| THERMOCOUPLE 7*THERMOCOUPLE 8 | 70.102 | 0.00000 |
| THERMOCOUPLE 3*THERMOCOUPLE 16 | 64.305 | 0.00000 |
| THERMOCOUPLE 6*THERMOCOUPLE 18 | 63.448 | 0.00000 |
| THERMOCOUPLE 6*THERMOCOUPLE 13 | 58.424 | 0.00000 |
| THERMOCOUPLE 2*THERMOCOUPLE 9 | 54.942 | 0.00000 |
| THERMOCOUPLE 3*THERMOCOUPLE 8 | 53.608 | 0.00000 |
| THERMOCOUPLE 7*THERMOCOUPLE 17 | 51.612 | 0.00000 |
| THERMOCOUPLE 4*THERMOCOUPLE 6 | 50.807 | 0.00000 |
| THERMOCOUPLE 11*THERMOCOUPLE 17 | 48.750 | 0.00000 |
| THERMOCOUPLE 3*THERMOCOUPLE 7 | 47.837 | 0.00000 |
| THERMOCOUPLE 1*THERMOCOUPLE 13 | 46.896 | 0.00000 |
| THERMOCOUPLE 6*THERMOCOUPLE 10 | 41.867 | 0.00000 |
| THERMOCOUPLE 1*THERMOCOUPLE 11 | 41.854 | 0.00000 |
| THERMOCOUPLE 7*THERMOCOUPLE 11 | 28.311 | 0.00000 |
| THERMOCOUPLE 4*THERMOCOUPLE 17 | 28.202 | 0.00000 |
| THERMOCOUPLE 1*THERMOCOUPLE 14 | 28.110 | 0.00000 |
| THERMOCOUPLE 6*THERMOCOUPLE 17 | 25.472 | 0.00000 |

| Source | LogWorth |  | PValue |
|---|---|---|---|
| THERMOCOUPLE 2*THERMOCOUPLE 3 | 22.883 |  | 0.00000 |
| THERMOCOUPLE 3*THERMOCOUPLE 4 | 21.724 |  | 0.00000 |
| THERMOCOUPLE 1*THERMOCOUPLE 3 | 21.043 |  | 0.00000 |
| THERMOCOUPLE 1*THERMOCOUPLE 7 | 18.172 |  | 0.00000 |
| THERMOCOUPLE 1*THERMOCOUPLE 2 | 17.611 |  | 0.00000 |
| THERMOCOUPLE 1*THERMOCOUPLE 16 | 16.956 |  | 0.00000 |
| THERMOCOUPLE 1 | 15.589 |  | 0.00000 ^ |
| THERMOCOUPLE 9*THERMOCOUPLE 12 | 14.589 |  | 0.00000 |
| THERMOCOUPLE 1*THERMOCOUPLE 18 | 13.723 |  | 0.00000 |
| THERMOCOUPLE 2*THERMOCOUPLE 6 | 12.253 |  | 0.00000 |
| THERMOCOUPLE 2*THERMOCOUPLE 4 | 10.585 |  | 0.00000 |
| THERMOCOUPLE 9*THERMOCOUPLE 17 | 5.845 |  | 0.00000 |
| THERMOCOUPLE 2*THERMOCOUPLE 7 | 4.709 |  | 0.00002 |
| THERMOCOUPLE 3*THERMOCOUPLE 13 | 2.903 |  | 0.00125 |
| THERMOCOUPLE 6*THERMOCOUPLE 11 | 1.746 |  | 0.01797 |